\documentclass[letterpaper,twocolumn,10pt]{article}
\usepackage{usenix}

\usepackage[english]{babel}
\usepackage{blindtext}

\usepackage{tikz}
\usepackage{amsmath}
\usepackage{amsfonts}
\usepackage{color}
\usepackage{xcolor}
\usepackage{xspace}
\usepackage{graphicx}
\usepackage{subcaption}
\usepackage{booktabs}
\usepackage[frozencache,cachedir=minted-cache]{minted}
\usepackage{enumitem}
\usepackage{totpages}
\usepackage[utf8]{inputenc}

\usepackage{filecontents}

\newcommand{\sysname}{{\textsc{CATO}}\xspace}
\newcommand{\sysnamebase}{{\textsc{CATObase}}\xspace}
\newcommand{\iotclass}{{\texttt{iot-class}}\xspace}
\newcommand{\appclass}{{\texttt{app-class}}\xspace}
\newcommand{\startupinf}{{\texttt{vid-start}}\xspace}
\newcommand{\all}{{\textsc{All}}\xspace}
\newcommand{\rfe}{{\textsc{Rfe10}}\xspace}
\newcommand{\mi}{{\textsc{Mi10}}\xspace}
\newcommand{\simanneal}{{\textsc{SimA}}\xspace}
\newcommand{\randsearch}{{\textsc{Rand}}\xspace}
\newcommand{\iterall}{{\textsc{IterAll}}\xspace}
\newcommand{\catonaivecost}{{\textsc{\sysname w/ na\"{i}ve cost}}\xspace}
\newcommand{\catonaiveperf}{{\textsc{\sysname w/ na\"{i}ve perf}}\xspace}
\newcommand{\catomodelinfcost}{{\textsc{\sysname w/ model inf cost}}\xspace}
\newcommand{\catopktdepthcost}{{\textsc{\sysname w/ pkt depth cost}}\xspace}
\newcommand{\myparab}[1]{\vspace{1pt}\noindent\textbf{#1}\hspace{0.5em}}

\begin{document}

\date{}

\title{CATO: End-to-End Optimization of ML-Based\\Traffic Analysis Pipelines}

\author{{\rm Gerry Wan}$^1$ \hspace{2px} {\rm Shinan Liu}$^2$  \hspace{2px} {\rm Francesco Bronzino}$^3$ \hspace{2px} {\rm Nick Feamster}$^2$ \hspace{2px} {\rm Zakir Durumeric}$^1$\\
$^1$Stanford University \hspace{2px} $^2$University of Chicago \hspace{2px} $^3$ENS Lyon}

\maketitle
\begin{abstract}
Machine learning has shown tremendous potential for improving the capabilities
of network traffic analysis applications, often outperforming simpler rule-based
heuristics. However, ML-based solutions remain difficult to deploy in practice.
Many existing approaches only optimize the predictive performance of their
models, overlooking the practical challenges of running them against network
traffic in real time. This is especially problematic in the domain of traffic
analysis, where the efficiency of the serving pipeline is a critical factor in
determining the usability of a model. In this work, we introduce CATO, a
framework that addresses this problem by jointly optimizing the predictive
performance and the associated systems costs of the serving pipeline. CATO
leverages recent advances in multi-objective Bayesian optimization to
efficiently identify Pareto-optimal configurations, and automatically compiles
end-to-end optimized serving pipelines that can be deployed in real networks.
Our evaluations show that compared to popular feature optimization techniques,
CATO can provide up to 3600$\times$ lower inference latency and 3.7$\times$
higher zero-loss throughput while simultaneously achieving better model
performance.
\end{abstract}


\section{Introduction}
Machine learning (ML) models have grown to outperform traditional rule-based
heuristics for a variety of traffic analysis applications, such as traffic
classification~\cite{kamath2015p4classification,liu2007kmeansclassification},
intrusion detection~\cite{wang2017malwareclassification}, and QoE
inference~\cite{bronzino2019inferring,mangla2019emimic}. Over the past few
years, researchers have explored various approaches to developing more accurate
models, ranging from better feature selection to employing sophisticated model
types and traffic
representations~\cite{moore2005discriminators,karagiannis2005blinc,fu2021whisper,piet2023ggfast,lotfollahi2020deep,
rimmer2017automated,zheng2022mtt,shapira2021flowpic,holland2021nprintml,abbasi2021dlsurvey,yang2021tfidf,boutaba2018mlsurvey}.
However, the predictive performance of ML-based solutions often overshadows an
equally critical aspect---the end-to-end efficiency of the serving pipeline that
processes network traffic and executes the model.

For traffic analysis, a significant challenge lies not just in developing
accurate models, but in meeting the performance demands of the network. Many
network applications must operate in real time with sub-second reaction times
and/or process hundreds of gigabits per second of traffic without packet
loss~\cite{wan2022retina}. Unfortunately, models developed without consideration
of the associated systems costs of serving them in real networks often turn out
to be unusable in practice~\cite{bronzino2021trafficrefinery}. Current
approaches to this problem typically rely on lightweight
models~\cite{liang2019neural}, programmable
hardware~\cite{xiong2019ilsy,kamath2015p4classification}, or early inference
techniques~\cite{bernaille2006earlyappid,peng2015packetnumber}, but many of
these unnecessarily compromise on predictive
performance~\cite{siracusano2022n3ic,bussegrawitz2019pforest,xiong2019ilsy}.

Recent studies have stressed the need to balance both the systems costs and
predictive performance of ML-based traffic analysis
solutions~\cite{bronzino2021trafficrefinery,siracusano2022n3ic}. However,
achieving this balance is difficult. The end-to-end latency and throughput of a
serving pipeline, which includes packet capture, feature extraction, \emph{and}
model inference, are difficult to approximate without in-network measurements.
Furthermore, the search space over optimal feature representations is
exponential in the number of candidate traffic features, and also depends on how
far into a flow to wait before making a prediction. The added complexity of not
just considering one objective, but two, makes end-to-end optimization of such
systems an open challenge.\looseness=-1

In this work, we present \sysname, a generalizable framework that systematically
optimizes the systems costs and model performance of ML-based traffic analysis
pipelines. We start by formalizing the development of ML models for traffic
analysis as a \emph{multi-objective} optimization problem. We then combine
multi-objective Bayesian optimization, tailored specifically for traffic
analysis, with a realistic pipeline profiler to efficiently construct end-to-end
optimized traffic analysis pipelines. \sysname simultaneously searches over the
selected features and the amount of captured traffic needed to compute those
features, factors which have been shown to significantly impact both efficiency
and predictive performance~\cite{bronzino2021trafficrefinery,jiang2023acdc}.
During this search, \sysname performs direct end-to-end measurements on the
resulting serving pipelines to both \emph{optimize} and \emph{validate} their
in-network performance.

We evaluate \sysname on live network traffic and offline traces across a range
of classification and regression traffic analysis tasks. Our experimental
results show that compared to popular feature optimization methods, \sysname can
reduce the end-to-end latency of the serving pipeline by up to 3600$\times$,
from several minutes to under 0.1~seconds, while simultaneously improving model
performance. Additionally, \sysname can increase zero-loss classification
throughput by up to 3.7$\times$.

We hope that our work helps to realize the potential impact of using machine
learning to manage and improve networks. Code is available at: {https://github.com/stanford-esrg/cato}.

\section{Background and Motivation}
\label{sec:motivation}

The networking community has long attempted to use machine learning (ML) to
perform traffic analysis tasks like QoE inference~\cite{bronzino2019inferring, mangla2019emimic, krishnamoorthi2017buffest,
gutterman2020requet}, traffic classification~\cite{sivanathan2019iotclass,
kamath2015p4classification, liu2007kmeansclassification}, intrusion
detection~\cite{wang2017malwareclassification}, and load
balancing~\cite{chang2023llb}. As traffic increasingly becomes encrypted, ML has
also been shown to be a promising technique for understanding otherwise opaque
network traffic, replacing traditional deep packet inspection and
other rule-based heuristics~\cite{singh2013mlsurvey,
nguyen2008mlsurvey,boutaba2018mlsurvey}.

While significant progress has been made in improving the predictive
capabilities of models used for traffic analysis~\cite{boutaba2018mlsurvey}, the
real-world deployability of a model is based not just on conventional notions of
ML performance (e.g., accuracy, F1 score), but also on the associated
systems-level performance (e.g., latency, throughput) of the entire serving
pipeline~\cite{bronzino2021trafficrefinery}. Given the real-time demands of
network operations, any applications that rely on ML must nonetheless operate
within tight performance budgets. Even small delays can cause substantial packet
loss and render a model
ineffective~\cite{bronzino2019inferring,hugon2023towards}, making systems
performance even more crucial for traffic analysis. As a result, ML-based
traffic analysis cannot solely target high predictive performance---the
end-to-end efficiency of the entire serving pipeline must be jointly optimized
as well.

\subsection{ML-Based Traffic Analysis} 
\begin{figure}
    \centering
    \includegraphics[width=\columnwidth]{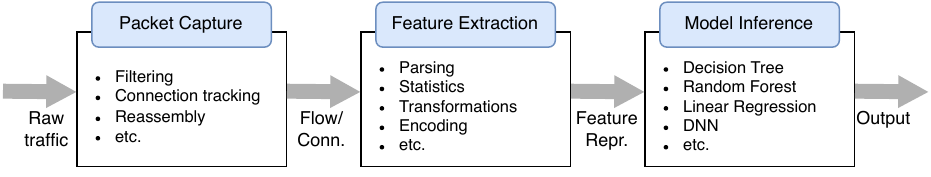}
    \caption{A typical serving pipeline for ML-based traffic analysis. Usability
	of a model hinges on both its predictive accuracy and the systems performance of the entire pipeline.}
    \label{fig:pipeline}
\end{figure}

ML-based traffic analysis typically
begins with the ingestion of raw traffic and ends with a prediction of a traffic
property, such as a service quality metric. While traffic analysis applications
are diverse, we focus on the class of problems that involves per-flow or
per-connection inference, such as traffic/device classification, QoE inference,
or intrusion detection. These applications typically make a prediction about an
entire flow or connection, then initiate an action such as triggering an alert,
blocking or rerouting the flow, or performing further analysis downstream
(Figure~\ref{fig:pipeline}).

Traffic inference extends beyond merely executing the model; it also involves
packet capture, connection tracking, flow reassembly, and feature extraction.
Raw traffic undergoes multiple operations, including header parsing,
computation, and encoding before arriving at the representation that is used as
input to the model. The final model inference step makes the prediction, with
its predictive performance determined by the model type (e.g., random forest,
neural network), the computed features, and the amount of data captured from the
flow. The end-to-end systems performance depends on \emph{all} of these aspects
together.

For traffic analysis in particular, the choice of features computed from the
network traffic is often as important, if not more so, than the model
itself~\cite{holland2021nprintml,jiang2023acdc,piet2023ggfast}. While many
previous works have focused on the model inference
stage~\cite{siracusano2022n3ic,lotfollahi2020deep,rimmer2017automated,zheng2022mtt,liang2019neural,xiong2019ilsy},
design decisions made in the earlier stages of the serving pipeline are crucial
to its performance and practicality, and warrant careful consideration.

\myparab{Optimizing Predictive Performance.} Many techniques have been proposed
to accurately make predictions about network traffic. These approaches range
from popular feature selection methods that choose highly predictive features
based on summary statistics~\cite{moore2005discriminators}, packet
lengths~\cite{bernaille2006earlyappid}, timing~\cite{karagiannis2005blinc},
and/or frequency domain characteristics~\cite{fu2021whisper}, to new techniques
like GGFAST~\cite{piet2023ggfast}, which generates specialized ``snippets'' to
classify encrypted flows.

Following advancements in domains like computer vision
and natural language processing, network researchers have also proposed
sophisticated deep learning models~\cite{lotfollahi2020deep,
rimmer2017automated,zheng2022mtt} and traffic
representations~\cite{shapira2021flowpic,holland2021nprintml} designed to
further optimize the predictive performance of traffic analysis applications.
However, most of these machine learning techniques are evaluated using offline
packet traces on metrics like accuracy, precision, recall, or F1 score,
overlooking the need to both \emph{optimize} and \emph{validate} their
in-network systems performance. As a result, ML-based traffic analysis solutions
that have been demonstrated to have high accuracy in controlled laboratory
experiments often turn out to be unusable in real-time deployments because of
the systems costs associated with running
them~\cite{bronzino2021trafficrefinery}.

\myparab{Optimizing Serving Efficiency.}
To address the systems requirements of traffic analysis on modern networks, some
works propose using lightweight models~\cite{liang2019neural,xiong2019ilsy}
or choose features that reduce model inference time~\cite{tong2014high}.
However, these
techniques can over-compromise on predictive performance for speed, and often overlook the efficiency of other pieces of the serving pipeline like packet
capture and feature extraction. Other approaches aim to optimize serving
efficiency by making predictions as early as
possible~\cite{bernaille2006earlyappid,bernaille2006onthefly,sena2009earlysvm,
dainotti2015earlymulticlass}. While many traffic analysis solutions implicitly
rely on the entire network flow or
connection~\cite{bronzino2019inferring,lee2020switchtree,shapira2021flowpic,barradas2021flowlens,aouini2022nfstream},
these ``early inference'' techniques make predictions after observing a
predefined number of packets. However, there is no ideal {\em packet depth}
(i.e., the number of packets to use from any given flow) that is universally
effective across applications. Choosing an appropriate value typically requires
prior domain-specific knowledge or resorting to manual
trial-and-error~\cite{peng2015packetnumber}. Consequently, existing works that
do explicitly choose a packet depth often opt for values such as 10, 50, or 100
packets with little justification~\cite{huang2013appr,piet2023ggfast,
holland2021nprintml,lopezmartin2017nniot,miettinen2017iotsentinel,rezaei2019deepappid,alan2016androidid,
siracusano2022n3ic}. As we will show in Section~\ref{sec:performance}, this
approach can miss significant opportunities for gains in both efficiency and model
performance.

\begin{figure}
    \centering
    \begin{subfigure}{0.49\columnwidth}
        \includegraphics[width=\linewidth]{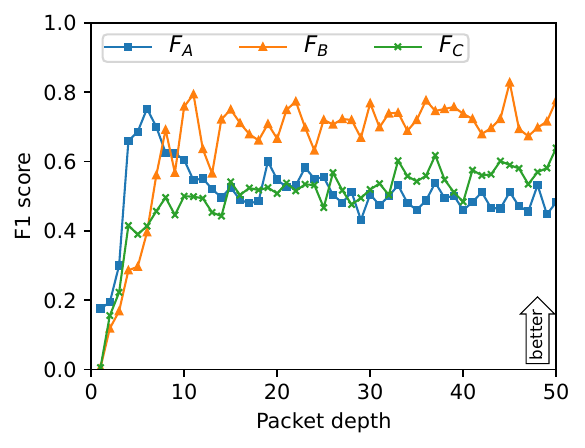}
        \caption{Packet Depth vs.\ F1 Score. The best feature sets differ at varying packet depths.}
        \label{fig:pktdepth_vs_perf}
    \end{subfigure}
    \hfill
    \begin{subfigure}{0.49\columnwidth}
        \includegraphics[width=\linewidth]{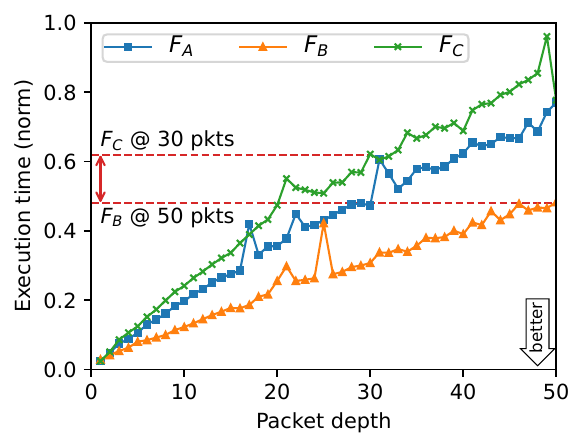}
        \caption{Packet Depth vs.\ Exec.\ Time. It can be cheaper to extract low-cost features at greater depths.}
        \label{fig:pktdepth_vs_cost}
    \end{subfigure}
    \caption{Effects of different (feature set, packet depth) configurations on F1 score and execution time. We highlight the size and complexity of the search space.}
    \label{fig:challenge}
\end{figure}

\subsection{Challenge of End-to-End Optimization}
\label{sec:joint_motivation}
Despite progress towards individually optimizing the predictive performance or
serving efficiency of ML-based traffic analysis, improvements to either area in
isolation often result in solutions that fail to achieve optimal combined model and
systems performance objectives. Systematically designing traffic analysis
pipelines that jointly optimize both of these objectives remains an open
challenge.

We illustrate this challenge by attempting to design an IoT device classifier
using the dataset published by Sivanathan et~al.~\cite{sivanathan2019iotclass}.
Our goal is to construct a serving pipeline that is \emph{Pareto-optimal} across
both its execution time and F1 score. In other words, it should not be possible
to further reduce the execution time without also reducing the F1 score, or vice
versa. We combine techniques from prior work by experimenting with different
features and early inference, both of which have been shown (and we
confirm) to significantly impact multiple aspects of the pipeline, including
predictive performance, model inference time, and feature extraction
time~\cite{jiang2023acdc,moore2005discriminators,peng2015packetnumber,bronzino2021trafficrefinery}.
For this example, we choose from six candidate flow features
(Appendix~\ref{app:candidate_features}, Table~\ref{tab:candidate_features}) and
vary the packet depth collected in each flow from 1 to 50, which is consistent
with values used in prior work~\cite{bernaille2006earlyappid,
peng2015packetnumber,piet2023ggfast,lopezmartin2017nniot}. We train the model,
compile the {complete} serving pipeline, and exhaustively measure the F1 score
and execution time for all $2^6 \times 50 = 3{,}200$ (feature set, packet depth)
combinations.

As seen in Figure~\ref{fig:challenge}, F1 score and execution time vary with the
chosen features and packet depth. For readability, we plot only three (labeled
$F_A/F_B/F_C$) out of the 64 possible feature sets, but find qualitatively
similar results across those not shown. We can see in
Figure~\ref{fig:pktdepth_vs_perf} that the best feature sets by F1 score differ
dramatically at different packet depths. While $F_A$ has the highest F1 score
within the first 10 packets, the ranking flips at higher packet counts.
Interestingly, the predictive performance of $F_B$ and $F_C$ increase with
packet depth, whereas the opposite is true for $F_A$. In
Figure~\ref{fig:pktdepth_vs_cost}, we observe that for the \emph{same} feature
set, execution time generally increases with packet depth. However, the overall
cost of waiting 50~packets to extract $F_B$ is lower than the cost of waiting
30~packets for $F_C$. This reveals that having the flexibility to optimize the
timing of feature extraction can significantly enhance serving efficiency, but
is not always as straightforward as simply minimizing analyzed packets. If
we look across both figures, we see that over-optimizing on execution time can
also adversely affect F1 score and vice versa. The non-linear trade-offs between
objectives further highlight the challenge in identifying Pareto-optimal
solutions without exhaustive measurement.

The trade-offs between predictive performance and systems costs form a
multi-dimensional and multi-objective search space that extends beyond merely
identifying which features result in more accurate models. It also includes
considerations for features that are efficient to extract, as well as how much
data must be captured to compute and represent the features. While exhaustive
measurement of end-to-end systems costs and model performance is feasible for
just six candidate features, it quickly becomes impractical when scaling up to
the dozens to hundreds of flow features typically considered by developers. In
our example, it took 5~days to train, compile, and measure all 3{,}200 serving
pipelines, but it would take over 7{,}000~years with 25~candidate features. The
size and complexity of the search space, coupled with the need to consider and
validate both model performance and systems cost objectives, makes end-to-end
optimization challenging. Addressing this challenge is the central contribution
of our work.

\section{Cost-Aware Traffic Analysis Optimization}
\label{sec:design}

We introduce \sysname (Cost-Aware Traffic Analysis Optimization), our solution
for cost-aware ML-based traffic analysis optimization. The goal of \sysname is
to automatically construct traffic analysis pipelines that jointly minimize the
end-to-end systems costs of model serving while maximizing predictive
performance. At its core, \sysname combines a multi-objective Bayesian
optimization-guided search with a novel pipeline generator and feature
representation profiler to produce serving pipelines suitable for deployment in
real networks.

\subsection{Problem Definition}
\label{sec:problem_defn}
\sysname takes as input a set of candidate network flow features, denoted by
$\mathcal{F}$. In line with conventional machine learning practice, these are
typically derived from domain expertise or determined by the capabilities of the
traffic collection tool. Common examples for traffic analysis include mean
packet size, bytes transferred, connection duration,
etc.~\cite{moore2005discriminators}, but can also include more complex features
like frequency domain ~\cite{yang2021noveltydetection,fu2021whisper} and
application-layer characteristics~\cite{bronzino2019inferring}. \sysname also
takes as input a maximum {connection depth} $N \in \mathbb{R}$, which serves as
an \emph{upper bound} on the amount of data in the connection that is considered
for inference. Concretely, this can be the number of initial packets (i.e.,
packet depth), bytes, or time into the connection prior to feature extraction.
These inputs define the \emph{search space} $\mathbb{X} =
\mathcal{P}(\mathcal{F}) \times {N}$, where $\mathcal{P}(\mathcal{F})$ is the
power set of $\mathcal{F}$, i.e., the set of all possible subsets of
$\mathcal{F}$, from which \sysname selects feature representations.

A \emph{feature representation} $x = (F,n)$ consists of a set of features $F
\subseteq \mathcal{F}$ and a value $n \leq N$ that indicates the connection
depth from which $F$ is extracted. Each feature representation gives rise to a
serving pipeline with an associated end-to-end systems cost and predictive
performance, denoted by the functions $\texttt{cost}(x): \mathbb{X} \rightarrow
\mathbb{R}$ and $\texttt{perf}(x): \mathbb{X} \rightarrow \mathbb{R}$. We note
that these functions are general and can be user-defined according to the
specific objectives of the traffic analysis problem. For instance,
$\texttt{cost}(x)$ can refer to the end-to-end inference latency, execution
time, (negative) throughput, etc., while $\texttt{perf}(x)$ can be defined as F1
score, accuracy, (negative) mean-squared-error, etc. We list a summary of
variables in Table~\ref{tab:notations}.

\begin{table}
    \centering
    \scriptsize
    \renewcommand{\arraystretch}{0.5}
    \begin{tabular}{ll}
    \toprule
    \textbf{Symbol} & \textbf{Description} \\
    \midrule
    $\mathcal{F}$ & Set of candidate network flow features \\
    $N$ & Maximum connection depth \\
    $\mathcal{P}(\mathcal{F})$ & Power set of $\mathcal{F}$ \\
    $\mathbb{X}$ & Search space defined as $\mathbb{X} = \mathcal{P}(\mathcal{F}) \times N$ \\
    $F$ & Set of features in a feature representation \\
    $n$ & Connection depth from which $F$ is extracted \\
    $x$ & Feature representation $x = (F, n)$\\
    $\texttt{cost}(x)$ & Systems cost objective function \\
    $\texttt{perf}(x)$ & Predictive performance objective function \\
    $\Gamma$ & Set of Pareto-optimal solutions\\
    \bottomrule
\end{tabular}
	\vspace{-5pt}
    \caption{Summary of Variables}
    \label{tab:notations}
\end{table}

\myparab{Multi-objective Optimization.}
We formalize the development of ML models for traffic analysis as a
{multi-objective} optimization over the search space $\mathbb{X}$. The aim is to
identify the Pareto front $\Gamma \subseteq \mathbb{X}$, which consists of all
non-dominated points in $\mathbb{X}$. In other
words, $\Gamma$ contains the maximally desirable feature set / connection depth
configurations, where no further improvement in systems cost or model
performance can be achieved without compromising the other objective.

We deliberately choose a multi-objective optimization over a single-objective
approach. Unlike a single-objective problem that aims to maximize model
performance while satisfying system constraints, or vice versa, a
multi-objective solution offers several advantages. The first is that the exact
system and model performance requirements may not be known a priori (e.g., due
to variable traffic rates or shared system resources), making it difficult to
precisely define constraints. Second, if requirements change, a single-objective
approach would necessitate redefining and rerunning the optimization with new
objectives and constraints~\cite{swamy2023homunculus}. By expressing the problem
as a multi-objective optimization, \sysname identifies multiple Pareto-optimal
solutions that each achieve a different balance between systems cost and model
performance, providing the flexibility to accommodate changing application needs
(e.g., adjusting accuracy thresholds or imposing new latency constraints)
without re-optimization.

\begin{figure}
    \centering
    \includegraphics[width=\columnwidth]{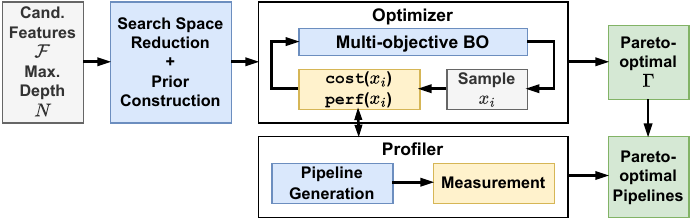}
    \caption{\sysname combines a multi-objective BO-based Optimizer and a realistic pipeline Profiler to construct and validate efficient ML-based traffic analysis serving pipelines.}
    \label{fig:design}
\end{figure}

\subsection{\sysname Overview}

\sysname constructs end-to-end optimized traffic analysis pipe-lines according
to the systems cost and model performance objective functions. It does so by
efficiently identifying Pareto-optimal feature representations, and generating
ready-to-deploy serving pipelines for a given model from those feature
representations. Figure~\ref{fig:design} depicts the high-level design, which
consists of the Optimizer and the Profiler:\looseness=-1

\begin{itemize}[noitemsep,topsep=0pt,leftmargin=12pt]

\item The Optimizer takes the set of candidate features and maximum connection
	depth, and performs a multi-objective Bayesian optimization-guided search
		over the feature representation space. It periodically queries the
		Profiler for the systems costs and model performance of its sampled
		feature representations, which it uses to further refine the search for
		the Pareto front. \looseness=-1

\item The Profiler accepts queries from the Optimizer, compiles binaries for the
	end-to-end serving pipeline, and runs them to accurately measure
		$\texttt{cost}(x)$ and $\texttt{perf}(x)$. These measurements serve the
		dual purpose of guiding the Optimizer towards Pareto-optimal solutions
		and validating the in-network performance of the generated pipelines.

\end{itemize}

\subsection{The \sysname Optimizer}
\label{sec:optimization}
In general, measuring $\texttt{cost}(x)$ and $\texttt{perf}(x)$ for an arbitrary
feature representation is computationally expensive. It involves generating the
serving pipeline, training and evaluating the ML model, and measuring
performance costs either through simulation or in physical testbeds. The massive
size of the search space and the computational cost of evaluating the objective
functions precludes the possibility of exhaustively searching all possible
configurations. To handle this intractability, \sysname leverages Bayesian
Optimization (BO), building on recent developments in multi-objective design
space exploration~\cite{nardi2019hypermapper} and sample-efficient
BO~\cite{hvarfner2022pibo} to efficiently estimate the Pareto front in
$\mathbb{X}$.

\myparab{Why Bayesian Optimization?}
\label{sec:bo}
Bayesian optimization is a technique designed for global optimization of
black-box objective functions~\cite{jones1998bo}, and has seen success in
domains like hyperparameter tuning~\cite{snoek2012mlbo,turner2021hpobo},
compiler optimization~\cite{nardi2019hypermapper,hellsten2023baco}, and
robotics~\cite{calandra2014robotbo}. It is particularly useful for
expensive-to-evaluate, non-linear objectives, as in our case with
$\texttt{cost}(x)$ and $\texttt{perf}(x)$. Moreover, the discontinuous nature of
these objective functions make the use of traditional gradient-based or linear
optimizers a poor fit for our problem.\looseness=-1

BO works by building a probabilistic surrogate model for the objective
function(s), and uses it to make decisions about which points (e.g., feature
representations) in the search space to evaluate next. Typically, BO begins by
sampling an initial number of points at random to build the surrogate model.
Each subsequent iteration involves generating a set of candidate points, using
the surrogate model to predict the objective function's output on the candidate
set, and choosing the next ``best''  (e.g., maximizing Expected Improvement)
point to evaluate using the real objective function. The surrogate model is then
updated to include the newly evaluated point, and the process repeats until some
stopping criteria is met, such as a maximum number of iterations or a
satisfactory level of convergence~\cite{jones1998bo, frazier2018botutorial}.
While BO is not guaranteed to work well in high-dimensional, multi-objective
spaces~\cite{shahriari2016boreview}, we describe how we augment it for our
context later in this section. We compare the performance of BO against other
search techniques in Section~\ref{sec:optimizer_eval} and show that our approach
is efficient at approximating the Pareto front.

\myparab{BO Formulation.} We formulate \sysname's search process as a
multi-objective Bayesian optimization problem with $|\mathcal{F}| + 1$
dimensions: one dimension per feature in $\mathcal{F}$ and one for the
connection depth $n$. Each feature parameter is represented by a binary
indicator variable, which denotes whether or not the feature is included. The
connection depth parameter is separately encoded as an integer or real-valued
variable upper bounded by the maximum connection depth $N$. This setup lets
\sysname \emph{concurrently} search over both features and connection depth
while optimizing for systems cost and model performance. We define this search
as a minimization of the two functions $\texttt{cost}(F,n)$ and
$-\texttt{perf}(F,n)$. These objective functions are managed by the Profiler
(Section~\ref{sec:profiling}), which generates complete serving pipelines
according to the feature representations sampled by the Optimizer, and returns
the end-to-end systems cost and model performance metrics.

\myparab{Tailoring BO for Traffic Analysis.}
\label{sec:preprocessing}
In its basic form, BO has several limitations. Conventional applications of BO
typically involve single-objective, low-dimensional (fewer than 20) search
spaces~\cite{frazier2018botutorial,shahriari2016boreview}. However, our traffic
analysis problem is inherently multi-objective, high-dimensional, and involves a
complex search space with mixed categorical (features) and numerical (connection
depth) variables. To address this, we augment the \sysname Optimizer with two
preprocessing techniques to improve its sample efficiency
(Figure~\ref{fig:design}). The first is a dimensionality reduction step that
strategically discards candidate features that are unlikely to improve the
model's predictive performance regardless of its impact on the end-to-end
systems costs. By default, we exclude features with a mutual
information~\cite{vergara2014mi} score of zero, which indicates no direct
informational relationship with the target variable.

The second technique incorporates prior probabilities into the BO formulation,
accelerating the search by providing the Optimizer with ``hints'' about the
approximate locations of Pareto-optimal feature representations. To account for
both objectives, \sysname constructs two sets of priors: one over the feature
space that targets $\texttt{perf}(x)$, and one over the connection depth that
targets $\texttt{cost}(x)$. The set of priors over the feature space encodes
each feature's relative contribution to the model's performance, and are derived
from the mutual information scores computed in the dimensionality reduction
step. Formally, we define the prior probability of whether a feature $f$ is part
of a Pareto-optimal feature representation $x = (F,n)$ as $\mathbb{P}(f \in F |
x \in \Gamma) =  (1 - \delta)\frac{I(f)}{I_{max}} + \frac{\delta}{2}$, where
$I(f)$ represents the mutual information of $f$ with respect to the target
variable, $I_{max}$ is the maximum mutual information among all candidate
features, and $\delta$ is a damping coefficient. The damping coefficient is used
to adjust the priors to prevent the feature with the highest mutual information
from always being included. $\delta=0$ signifies
no damping, while $\delta=1$ results in uniform priors for all features. These
probabilities encourage \sysname to more frequently explore regions of the
search space that include features with higher predictive power.

The prior over the connection depth is represented by a probability mass
function that decays linearly as the connection depth increases. The rationale
is that for the same feature set, waiting longer to capture more packets or
bytes before feature extraction correlates with worse systems performance. This
prior encourages \sysname to more frequently explore representations that
require fewer packets or less data, despite not requiring domain-specific
knowledge about the {optimal} connection depth at which to collect features. As
we will show in Section~\ref{sec:microbenchmarks}, \sysname is robust to
reasonably large connection depth ranges.

We emphasize that these preprocessing steps can be performed efficiently without
needing to evaluate the objective functions. Despite the term ``prior,'' no
prior knowledge about the optimal features or connection depth needs to be
supplied by the user. \sysname automatically derives the priors used to
accelerate its search, thereby streamlining its usability.

\subsection{The \sysname Profiler}
\label{sec:profiling}
The \sysname Profiler evaluates the feature representations sampled by the
Optimizer based on the concrete definitions of $\texttt{cost}(x)$ and
$\texttt{perf}(x)$. To accomplish this, it generates code for the
packet capture and feature extraction stages of each sampled point, trains the
model, and runs the full serving pipeline to \emph{directly measure} its
end-to-end systems costs and model performance. This measurement serves two
purposes: (1) guiding the search process of the Optimizer, and (2) validating
the in-network performance of identified solutions. 

\myparab{Why Measure?} Using heuristics to estimate the end-to-end systems cost
of a traffic analysis pipeline is difficult. Much like how existing heuristics
that approximate model performance often fail to capture interdependencies and
correlations among features~\cite{vergara2014mi,altmann2010pimp,jiang2023acdc},
systems cost heuristics can similarly fail to capture the complexities of
packet capture and feature extraction. Traffic analysis
is particularly sensitive to this, since the processing steps during feature
extraction often overlap in non-trivial ways. For instance, computing mean TCP window size and the number of ACKs sent in a connection require
parsing each packet down to its TCP header, a shared task that must be factored
into the end-to-end cost. Likewise, computing the mean window size also involves
calculating its sum, the latter of which can then essentially be used for free.
Other factors like resource contention and characteristics of the network
traffic (e.g., bursty vs. non-bursty) can also unpredictably affect the
end-to-end systems cost~\cite{romero2021infaas}. 

We argue that rather than trying to model or predict these complex systems-level
interactions, it is both more accurate and useful to perform direct measurement.
With direct measurement, \sysname captures the actual end-to-end cost of the
serving pipeline, encompassing all critical components including packet capture,
feature extraction, and the model inference itself. Accurate measurement not
only helps the Optimizer make well-informed decisions, but also helps users
build confidence in validating whether identified solutions are operationally
viable. Although this approach can be computationally expensive, the cost of
training the model, generating the full serving pipeline, and measuring its
performance is balanced by the sample efficiency of the Optimizer.

\myparab{Pipeline Generation.}
To evaluate different feature representations during the search process, we
require an \emph{automated} way to measure $\texttt{cost}(x)$ and
$\texttt{perf}(x)$ for any $x \in \mathbb{X}$. With a search space size of
$O(2^{|\mathcal{F}|} \times N)$, manually implementing packet capture,
feature extraction, and model inference for each evaluated point is
impractical. One approach that enables flexible evaluation is ``runtime
branching,'' which uses branching logic at runtime to determine which paths in
the code should be executed to extract a given feature representation. However, runtime branching introduces additional overhead that can
contaminate the cost measurements of performance-sensitive traffic analysis
pipelines. Instead, \sysname employs \emph{conditional compilation} to build and
run customized end-to-end serving pipelines tailored to each configuration. The
resulting binary matches the performance of a manually implemented pipeline,
containing only the set of operations needed to collect traffic data up to the
specified connection depth, extract the corresponding features, and execute the
model inference. This technique not only constructs fully operational traffic
analysis pipelines, but also provides the flexibility to accurately measure any
point in the search space.

\myparab{Pipeline Measurement.}
\sysname presents a testbed interface that replicates a real-world deployment
scenario of the pipeline. For model performance measurements, the Profiler
trains a fresh model for each representation sampled by the Optimizer and
directly measures its predictive performance to account for any interaction
effects between features. The final performance metric is derived from a
hold-out test set. We note that \sysname operates on pre-labeled datasets,
meaning it does not focus on automatic labeling or ground truth generation. The
framework is designed to optimize model accuracy and system performance based on
this labeled input. For systems cost measurements, \sysname either simulates
traffic inputs from the training data, or, when feasible, deploys the full
serving pipeline in its target network environment (e.g., a passive monitoring
or bump-in-the-wire deployment model) for end-to-end measurements. While each
measurement can be expensive, the Optimizer is intentionally designed to minimize
the number of measurements needed to approximate the Pareto front. We report
wall-clock times for several of our evaluated use cases
(Section~\ref{sec:datasets}) in Appendix~\ref{app:wallclock}.

\section{Implementation of \sysname}
\label{sec:implementation}

We detail our implementation of \sysname, covering the Optimizer, Profiler,
model training, and objective functions.

\myparab{Bayesian Optimization.}
We implement the \sysname Optimizer using
HyperMapper~\cite{nardi2019hypermapper}, a Bayesian optimization framework for
design space exploration. HyperMapper supports multi-objective optimization with
mixed-variable search spaces, but is not tailored specifically for
high-dimensional BO. We use $\pi$BO~\cite{hvarfner2022pibo} for prior injection,
but adapt its implementation to incorporate \sysname-generated priors in
multi-objective use cases. We use a random forest as the surrogate model, which
has been shown to perform well compared to more traditional Gaussian processes
for discontinuous and non-linear objective
functions~\cite{nardi2019hypermapper}. The prior over the packet depth is
constructed using the Beta distribution with $\alpha=1$ and $\beta=2$. We
initialize the Optimizer with three iterations of random search space
exploration and choose $\delta = 0.4$ based on empirically tuned values
(Section~\ref{sec:microbenchmarks}).

\myparab{Pipeline Generation.}
The \sysname Profiler generates serving pipelines using a modified version of
Retina~\cite{wan2022retina}, a Rust framework that compiles traffic
\emph{subscriptions} into efficient packet processing pipelines. A subscription
defines the rules for how incoming traffic should transformed into a specific
representation, and invokes a callback on the returned data. We implement the
model inference stage in the callback, and subscribe to a template feature
representation that can be modified at compile-time to the specific
representation being evaluated. To dynamically generate the custom packet
capture and feature extraction stages, we create a Retina subscription module
that implements the processing steps needed to extract \emph{all} candidate
features. Each operation (e.g., parse an IPv4 header, add to a cumulative sum of
packet inter-arrival times, etc.) is then annotated with a configuration
predicate that specifies the subset of features necessitating its execution. If
the feature representation being evaluated contains at least one of the
predicated features, the predicate evaluates to true and the operation is
conditionally compiled into the binary. This technique avoids redundant
computation in shared steps, such as parsing headers, and
ignores operations associated with features that are not included. For packet
capture, we annotate the subscription with an early termination flag that stops
data collection once the connection depth is reached.

\begin{figure}
    \centering
    \scriptsize
    \input{figures/conditional_compilation}
    \caption{An example portion of the \sysname Profiler's template subscription
    module. Each operation is predicated on its associated features and
    conditionally compiled with the $\texttt{cfg}$ macro. For instance, if the
    evaluated feature set consists of $\texttt{ttl\_min}$ and
    $\texttt{winsize\_max}$, then only lines 9, 11, 13, 16, and 17 will execute
    on each new packet, and only those two features will be extracted in the
    final feature representation. This enables dynamic cost profiling that
    matches the characteristics of a manually implemented feature extraction
    stage.}
    \label{fig:conditional_compilation}
\end{figure}

Figure~\ref{fig:conditional_compilation} shows pseudo-code for an
example portion of the template subscription module. We implement 67~candidate
features (Appendix~\ref{app:candidate_features},
Table~\ref{tab:candidate_features}) in $1{,}600$ lines of Rust code. We note
that the chosen candidate features are not specific to any use case: they are
widely used in traffic analysis applications and are common
features exposed by open source
tools~\cite{moore2005discriminators,zeek,wan2022retina,bussegrawitz2019pforest,siracusano2022n3ic,bronzino2019inferring,yang2021noveltydetection,drapergil2016vpndetection}.
We use number of packets into the connection to measure connection
depth.

We note that our Profiler implementation uses Retina~\cite{wan2022retina} to
target commodity servers. However, \sysname's core design principles remain
applicable to optimizing and validating hardware-based traffic analysis pipelines, which we discuss further in Section~\ref{sec:related}.

\myparab{Model Training.}
As an optimization framework, \sysname is general to the specific type of model
used in the traffic analysis pipeline. We implement support for three model
types: decision trees (DT), random forests (RF), and deep neural networks (DNN).
For DT and RF, we use scikit-learn's DecisionTreeClassifier and
RandomForestClassifier, with 5-fold nested cross validation and grid search for
hyperparameter tuning. We tune the maximum tree depth from 3--20 and set the
number of estimators to 100 for RF. To match the speed of the Rust-based feature
extraction stage, we retrain the best-performing DT and RF models in Rust using
the SmartCore~\cite{smartcore} library and evaluate the final Rust model on a
hold-out test set containing 20\% of the data. 

For DNN, we implement a fully connected feedforward neural network in
TensorFlow, consisting of three hidden layers with ReLU activation and L2
regularization. We apply dropout to prevent overfitting and use the Adam
optimizer for training. Since Rust lacks mature DNN libraries, we train and
evaluate the DNN models entirely in Python/TensorFlow. Additional details are
provided in Appendix~\ref{app:model_details}.

\myparab{Objective Functions.}
We use end-to-end inference latency, zero-loss classification throughput, and
pipeline execution time as three different metrics for systems cost. End-to-end
inference latency measures the duration from the arrival of the first packet in
the connection to the model's final prediction. This includes the time spent
extracting features from raw traffic, the model inference time, and time spent
waiting for packets to arrive. Zero-loss throughput is the highest ingress
traffic rate that can be sustained by the serving pipeline with no packet drops,
which we negate to match the sign of $\texttt{cost}(x)$ minimization. The
execution time measures the total CPU time spent in the serving pipeline,
excluding time between packets. This metric is less dependent on the specific
characteristics of the input traffic, and is an indirect measure of both latency
and throughput. Although these can be combined into a single cost metric, we
evaluate them separately to show \sysname's flexibility. Depending on the
traffic analysis use case, which we detail in the next section, we use either
the F1 score or root-mean-squared error, calculated from the predictions on the
hold-out test set as the model performance metric.

\section{Evaluation}
\label{sec:evaluation}
We evaluate \sysname over a variety of configurations and use cases. Section~\ref{sec:datasets} details our datasets and testbeds. In
Section~\ref{sec:performance}, we show that \sysname can help traffic analysis
applications achieve substantially lower inference latency and higher throughput
without compromising model performance, and in many cases improve upon both
metrics. We also compare it with Traffic Refinery~\cite{bronzino2021trafficrefinery}, a recent system for cost-aware ML on network traffic. Section~\ref{sec:optimizer_eval} compares the efficiency of the
\sysname Optimizer to alternative Pareto-finding approaches, and
Section~\ref{sec:profiler_eval} performs an ablation study of the Profiler. In
Section~\ref{sec:microbenchmarks}, we run micro-benchmarks on \sysname's
sensitivity to various search space sizes and hyperparameters.

\begin{table}
    \centering
    \footnotesize
    \renewcommand{\arraystretch}{0.6}
    \begin{tabular}{llll}
    \toprule
    Use Case    & Type           & Traffic & Model \\
    \midrule
    \appclass   & Classification & Live    & Decision Tree    \\
    \iotclass   & Classification & Dataset & Random Forest    \\
    \startupinf & Regression     & Dataset & Deep Neural Network   \\
    \bottomrule
\end{tabular}
	\vspace{-5pt}
    \caption{Evaluation Use Cases}
    \label{tab:use_cases}
\end{table}

\subsection{Datasets and Testbeds}
\label{sec:datasets}

We consider three use cases in our evaluations: web application classification
(\appclass), IoT device recognition (\iotclass), and video startup delay
inference (\startupinf). These are typical analysis tasks of varying complexity
that are representative of the type of ML-based inference performed on network
traffic. Table~\ref{tab:use_cases} summarizes them, with more details provided
in Appendix~\ref{app:datacollection_details}.

\myparab{Web Application Classification.} While open-source traffic
classification datasets exist, replaying them at modern line rates is challenging without duplicating flows. To evaluate
serving pipelines against real traffic at high speeds, we develop a use case
that identifies one of six common web applications from live traffic on a large
university network. This type of classification is typically used by web
application firewalls or in the early stages of network QoE inference
pipelines~\cite{bronzino2019inferring,azurewaf}. For ground truth, we label
connections using the server name in the TLS handshake. We train and evaluate
decision tree models using flow statistics captured from the network, then
deploy them to the same network for real-time serving using
Retina~\cite{wan2022retina}.

\myparab{IoT Device Recognition.} To help make our results reproducible, we also
consider an IoT device recognition use case based on the dataset published by
Sivanathan et~al.~\cite{sivanathan2019iotclass}. We use a random forest to
classify connections as belonging to one of 28~IoT device types. Although
real-time throughput experiments are not feasible without duplicating flows, we
use this dataset to report micro-benchmarks and evaluate \sysname's ability to
approximate the true Pareto front.

\myparab{Video Startup Delay Inference.} We further demonstrate \sysname's
generalizability to different traffic analysis tasks and model types through a
regression use case that predicts the startup delay of video streams. Startup
delay inference is widely used in analysis of encrypted video traffic as a
measure of QoE~\cite{bronzino2019inferring,mazhar2018qoe,balachandran2013qoe}.
We choose startup delay (rather than other QoE metrics) to provide a regression
task that complements the previous two classification use cases. We also adopt a
more complex DNN instead of a tree-based model, using the YouTube dataset
published by Bronzino et al.~\cite{bronzino2019inferring}.

\begin{figure*}[t]
    \centering
    \begin{subfigure}[t]{0.24\linewidth}
        \centering
        \includegraphics[width=\linewidth]{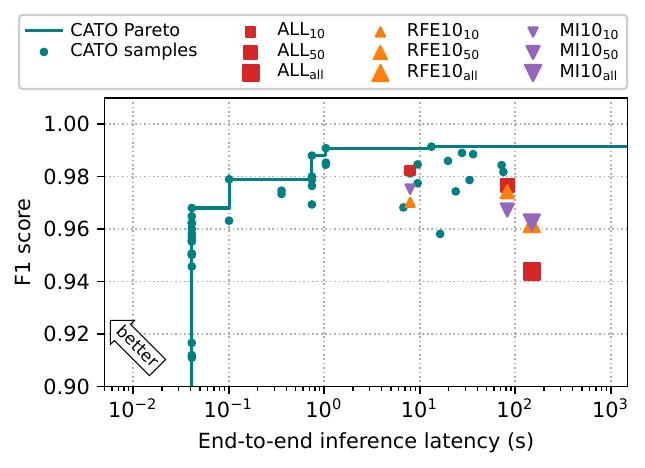}
        \caption{\iotclass latency}
        \label{fig:compare_latency_iot}
    \end{subfigure}
    \hfill
    \begin{subfigure}[t]{0.24\linewidth}
        \centering
        \includegraphics[width=\linewidth]{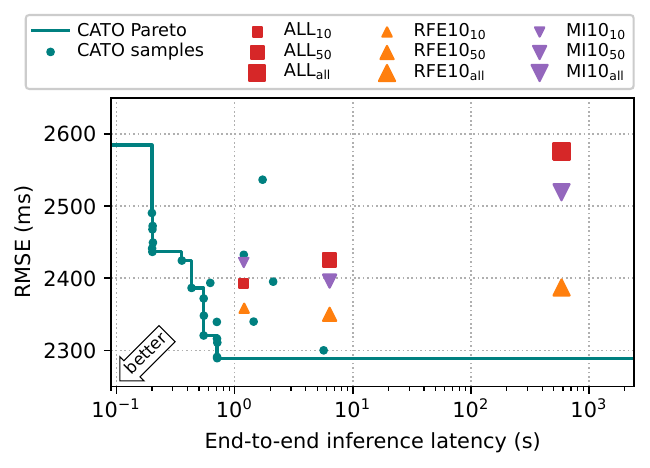}
        \caption{\startupinf latency}
        \label{fig:compare_latency_video}
    \end{subfigure}
    \hfill
    \begin{subfigure}[t]{0.24\linewidth}
        \centering
        \includegraphics[width=\linewidth]{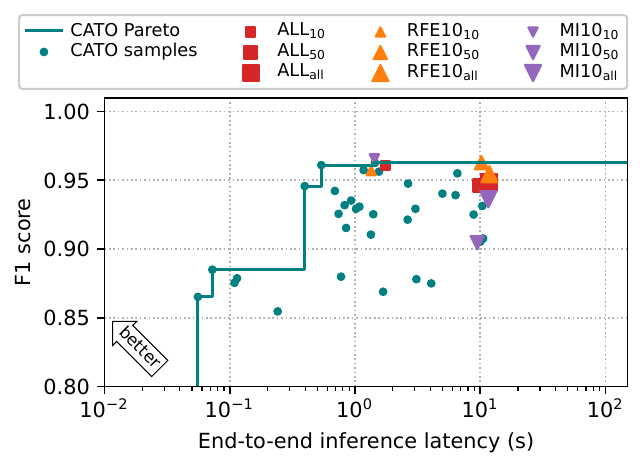}
        \caption{\appclass latency}
        \label{fig:compare_latency_app}
    \end{subfigure}
    \hfill
    \begin{subfigure}[t]{0.24\linewidth}
        \centering
        \includegraphics[width=\linewidth]{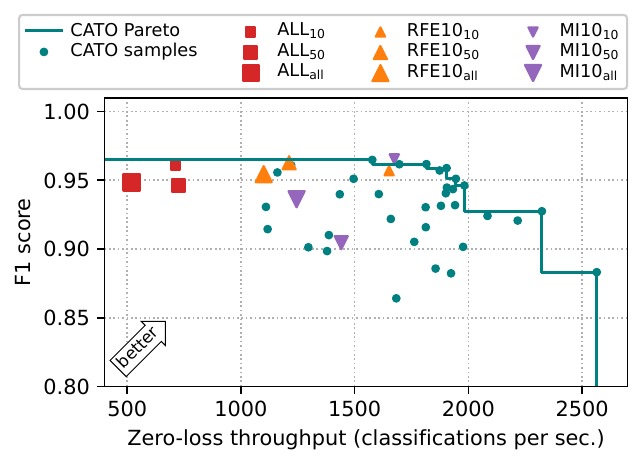}
        \caption{\appclass throughput}
        \label{fig:compare_throughput_app}
    \end{subfigure}

    \caption{Comparison of F1 score / RMSE vs. end-to-end inference latency /
	zero-loss throughput (single-core) for \iotclass, \startupinf, and \appclass
	serving pipelines. \sysname identifies multiple solutions on its Pareto
	front that dominate those found by traditional optimization techniques, and
	can achieve significantly better systems and predictive performance. }
\end{figure*}

\subsection{Model Serving Performance}
\label{sec:performance}
We first examine the end-to-end inference latency, zero-loss throughput, and
predictive performance of \sysname-optimized serving pipelines. Note that
\sysname itself is not a classifier, but a general framework for optimizing
ML-based serving pipelines for real-time traffic analysis. Therefore, instead of
directly comparing \sysname with existing classifiers or models, we evaluate it
against optimization strategies commonly used in prior work to build those
models. We use the following feature optimization methods and combine them with
early inference techniques as our baselines:
\begin{itemize}[noitemsep,topsep=0pt,leftmargin=15pt]
    \item \all: Use all available features.
    \item \rfe: Select the top ten features by recursive feature
    elimination~\cite{guyon2002rfe}. RFE trains a model using all available
    features, then iteratively removes the least important feature and retrains
    until the desired number remains.
    \item \mi: Select the top ten features based on mutual
    information~\cite{vergara2014mi}. This is a model agnostic algorithm that
    measures how much information each feature contributes to the target
    variable and picks the most relevant ones.
\end{itemize}
Prior traffic analysis solutions typically wait until the end of the connection
before making a
prediction~\cite{bronzino2019inferring,lee2020switchtree,shapira2021flowpic,barradas2021flowlens,aouini2022nfstream}
or use a fixed packet depth for early
inference~\cite{holland2021nprintml,lopezmartin2017nniot,bernaille2006onthefly,bernaille2006earlyappid,miettinen2017iotsentinel,rezaei2019deepappid,alan2016androidid}.
For example, Peng et~al.~\cite{peng2015packetnumber} collects up to the first 10
(including TCP handshake) packets, while recent work like
GGFAST~\cite{piet2023ggfast} use the first 50. For a thorough analysis, we
compare against these strategies by running each baseline at packet depths of
10, 50, and all packets. \sysname does not assume a predefined optimal packet
depth, but searches over the entire feature representation space as part of its
optimization process. We choose a maximum packet depth of 50 and run for
50~iterations, which is consistent with common machine learning
practices~\cite{bouthiller2020mlsurvey,hvarfner2022pibo}. We show how \sysname
reacts to different packet depth ranges in Section~\ref{sec:microbenchmarks}.

\myparab{End-to-End Inference Latency.}
Figures~\ref{fig:compare_latency_iot},~\ref{fig:compare_latency_video},
and~\ref{fig:compare_latency_app} show the end-to-end inference latency and
predictive performance (F1 score or RMSE) for \iotclass, \startupinf, and
\appclass, respectively. Each \sysname sample represents a candidate point
explored during the optimization process, with \sysname's Pareto front
constructed from the set of non-dominated points. For \iotclass and \startupinf,
all points on \sysname's Pareto front dominate the baseline solutions, achieving
equal or better predictive performance with lower end-to-end latency. For
\iotclass, \sysname can reduce the inference latency by 11--79$\times$ compared
to solutions that use the first 10~packets in the connection, 817--2000$\times$
compared to those that use the first 50, and over 3600$\times$ (from several
minutes to under 0.1~seconds) compared to those that wait until the end of the
connection. Likewise for \startupinf, \sysname generates solutions that can
infer video startup delays in less than one second (a 2.2--2900$\times$ speedup,
depending on baseline) while also reducing the mean squared error of its
predictions.

Since end-to-end inference latency is largely dominated by packet inter-arrival
times, this improvement can be attributed to \sysname's ability to find
alternative sets of features using the fewest packets necessary without
compromising the predictive performance of these popular feature selection
methods. For example, \rfe using the first 10~packets ($\rfe_{10}$) in \iotclass achieves an
F1 score of 0.970 with an inference latency of 7.9~seconds. However, \sysname
identifies a different set of features using just the first 3~packets for a
better F1 score of 0.979 and an inference latency of 0.1~seconds.

We find a similar pattern for \appclass, where \sysname-optimized pipelines
outperform most baseline methods across both objectives. While $\mi_{10}$ and $\rfe_{50}$
achieve slightly higher F1 scores (0.963 and 0.962), \sysname produces a
solution with a nearly identical F1 score (0.960) and a latency of
0.54~seconds---2.6$\times$ and 19$\times$ faster than $\mi_{10}$ and
$\rfe_{50}$, respectively. These results reinforce that for traditional feature
optimization methods, it is not always clear a priori which feature set at which
packet depth results in the best model performance or serving efficiency.
Through end-to-end optimization of both objectives over the entire feature
representation space, \sysname is able to automatically derive and validate the
performance of faster and more accurate traffic analysis pipelines.

\myparab{Zero-Loss Classification Throughput.}
We compare the predictive performance and classification throughput of solutions
found by \sysname with those found by the baseline methods for \appclass. We
exclude \iotclass and \startupinf due to limitations in replaying the traces at
high speeds without repeating flows. For a realistic assessment, we use live
traffic from our campus network, but restrict all experiments to a
{single} core to avoid saturating our network's maximum ingress
throughput. In an actual deployment scenario, the throughput can be easily
scaled up by adding more cores, owing to the per-core scalability of
Retina~\cite{wan2022retina}. More details about our throughput experiments can
be found in Appendix~\ref{app:measurement_details}.

Figure~\ref{fig:compare_throughput_app} shows that \sysname's solutions
outperform the baselines in both throughput and F1 score, with the
exception of $\mi_{10}$. Despite this, \sysname successfully
identifies the feature representation with the highest overall F1 score
and the one with the highest zero-loss throughput. For a decrease in F1 score
from 0.96 to 0.93, \sysname can increase throughput by 37\%. Compared to
solutions that wait until the end of the connection, \sysname can improve the
zero-loss throughput by a factor of 1.6--3.7$\times$, and 1.3--2.7$\times$ for
those that require the first 50~packets while also achieving higher model
performance. Notably, \sysname achieves these results after
exploring just 50~feature representations out of $2^{67} \times 50 = 7 \times
10^{21}$ (67~candidate features, up to a maximum packet depth of 50).

\myparab{Comparison with Traffic Refinery.}
We further compare \sysname with Traffic
Refinery~\cite{bronzino2021trafficrefinery}, a recent traffic analysis framework
that also facilitates joint evaluation of model performance and system costs.
Unlike \sysname, Traffic Refinery requires {manual} exploration of flow features
and connection depth. We simulate Traffic Refinery's cost profiler using
\sysname's execution time cost metric and replicate its built-in packet counter
(\textsc{PC}), packet timing (\textsc{PT}), and TCP counter (\textsc{TC})
feature classes. While Traffic Refinery defaults to making an inference after
ten seconds into the flow, we evaluate it at packet depths of 10, 50, and
all packets for consistency with the above baselines (see
Appendix~\ref{app:traffic_refinery} for more details).

\begin{figure}
    \centering
    \includegraphics[width=\columnwidth]{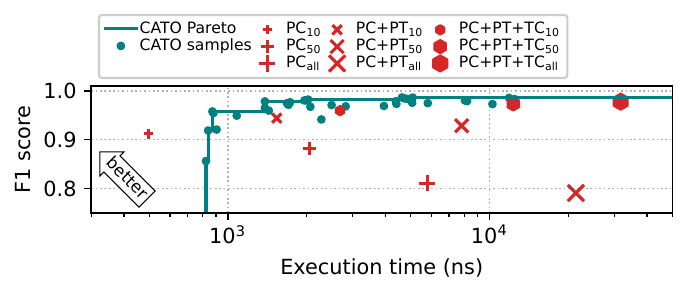}
	\vspace{-20pt}
    \caption{F1 score vs. pipeline execution time for \iotclass using \sysname and Traffic Refinery (in red). While $\textsc{PC}_{10}$ achieves a strong trade-off due to its low cost, \sysname more consistently finds solutions closer to the Pareto front.}
    \label{fig:tr_compare}
\end{figure}

Figure~\ref{fig:tr_compare} plots F1 score vs. execution time for \iotclass
using \sysname and Traffic Refinery. The results show that Traffic Refinery’s
macro-aggregation of standard feature classes is less efficient than \sysname at
finding optimal trade-offs. \sysname achieves better accuracy at lower cost
across all sampled points except $\textsc{PC}_{10}$. Even in this case, it is
still able to identify an alternative representation with a similar F1 score and
a modest 344~ns increase in pipeline execution time. While packet counters alone
at a packet depth of 10 happen to perform well for \iotclass at minimal cost,
adding packet timing and TCP state information further improves accuracy but
incurs a much higher execution time. Identifying such configurations with
Traffic Refinery requires trial-and-error, as it is not clear which combinations
perform best at which packet depths.
In contrast, \sysname more efficiently explores the feature representation
space, clustering solutions closer to the Pareto front and more effectively
balancing predictive performance and systems cost across a broader range of
configurations.

\subsection{Optimizer Efficiency}
\label{sec:optimizer_eval}
In this section, we evaluate the efficiency of the \sysname Optimizer by
measuring the quality of its computed Pareto front and the speed at which it
converges to the true Pareto front. Since it is infeasible to exhaustively
measure all points in~$\mathbb{X}$ to obtain the true Pareto front for the
default candidate feature set size of 67, we evaluate on a smaller candidate set
of six features (Appendix~\ref{app:candidate_features},
Table~\ref{tab:candidate_features}) for the \iotclass use case in order to
obtain the ground truth. We compare the Optimizer with three alternative
Pareto-finding algorithms:
\begin{itemize}[noitemsep,topsep=0pt,leftmargin=15pt]
    \item \simanneal: Use simulated annealing~\cite{kirkpatrick1983anneal}, a
    metaheuristic for solving optimization problems with a complex search space.
    We provide details of our multi-objective implementation in
    Appendix~\ref{app:simanneal_impl}.
    \item \randsearch: Sample a random subset of features at a random packet
    depth without replacement.
    \item \iterall: Use all available features but increment the packet depth on
    each iteration, starting from one.
\end{itemize}
We remark that these algorithms differ from the baseline methods described in
the previous section in that they attempt to estimate the Pareto front rather
than a single point solution. In $N$~search iterations, each of these
alternative approaches makes exactly $N$~calls to $\texttt{cost}(x)$ and
$\texttt{perf}(x)$.

\myparab{Pareto Front Quality.} We compare the quality of the Pareto front
estimated by each algorithm using Hypervolume Indicator (HVI). HVI is a common
metric used to compare multi-objective algorithm performance, and measures the
area between the estimated Pareto front and the true Pareto front bounded by a
reference point~\cite{li2019qualityeval}. We use pipeline execution time as our
chosen systems cost metric and F1 score for model performance. Since F1 score
and execution time have different raw value ranges, we normalize the data before
computing HVI to assign similar importance to both objectives.

\begin{figure}
    \centering
    \includegraphics[width=\columnwidth]{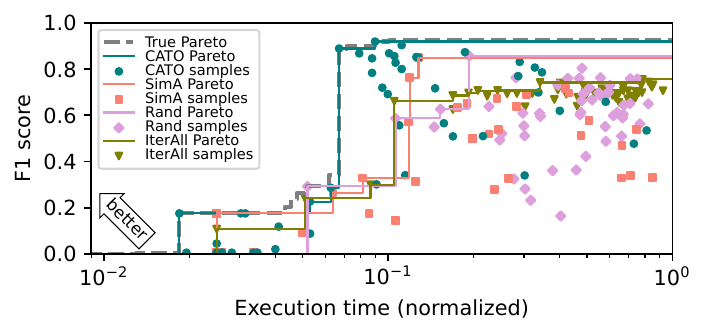}
    \vspace{-18pt}
    \caption{Estimated Pareto fronts after 50~iterations. \sysname outperforms other Pareto-finding approaches, especially in regions with high F1 scores and low execution times. The sampled points illustrate the candidate solutions explored by each algorithm during the optimization process, while only the Pareto-optimal points are included in the final output.}
    \label{fig:pareto_quality}
\end{figure}

We run each search algorithm for 50~iterations at a maximum packet depth of 50.
In Figure~\ref{fig:pareto_quality}, we plot the feature representations sampled
at each iteration alongside the corresponding Pareto front constructed from the
non-dominated points. These sampled points represent candidate solutions
explored during optimization and do not directly impact the end-to-end traffic
analysis pipeline. While many intermediate samples overlap between algorithms
due to inherent randomness, they primarily serve to illustrate the regions
explored by each algorithm and are not included in the final solution. For
reference, we also show the true Pareto front computed from exhaustively
measuring all $2^6 \times 50 = 3{,}200$ feature representations in the search
space.

We observe that \sysname's Pareto front closely approximates the true Pareto
front while sampling less than 1.6\% of the search space. Using a worst-case
reference point (F1 score of 0 and normalized execution time of 1), \sysname
achieves an HVI of 0.98, compared to 0.88, 0.86, and 0.77 achieved by
\simanneal, \randsearch, \iterall, respectively. We note that \sysname does not
entirely dominate all alternative search algorithms, especially around F1 scores
of 0.3. However, it performs better at higher and lower extremes. This behavior
can be attributed to \sysname's tendency to explore representations that require
very few packets (due to the decay-shaped prior placed over connection depth),
while also injecting priors based on each feature's relative importance. If we
only consider solutions with an F1 score of at least 0.8, the HVI is 0.95 for
\sysname, 0.39 for \simanneal and \randsearch, and 0 (no solutions found) for
\iterall.

\begin{figure}
    \centering
    \includegraphics[width=\columnwidth]{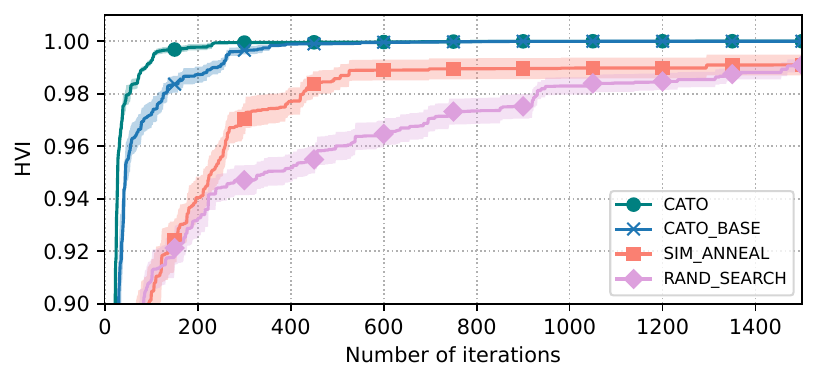}
    \vspace{-12pt}
    \caption{\sysname efficiently converges to the true Pareto front. We show
    the mean and standard error of the HVI with a worst-case reference point
    across 20~runs.}
    \label{fig:convergence}
\end{figure}

\myparab{Convergence Speed.}
Figure~\ref{fig:convergence} compares the sample efficiency of \sysname with
alternative search algorithms. We extend the 50~sample explorations commonly
used by ML practitioners~\cite{bouthiller2020mlsurvey,hvarfner2022pibo} to
$1{,}500$ to examine the convergence rate towards the true Pareto front as most of
the search space is explored. We also plot the performance of \sysname's
baseline BO formulation without dimensionality reduction and prior injection
(\sysnamebase). We exclude \iterall from this analysis since more than
50~iterations would exceed the maximum packet depth covered by the ground truth
Pareto front. Moreover, we find that the HVI for \iterall does not show
significant improvement beyond this point.

We can see that \sysname converges to the true Pareto front fastest,
demonstrating a more sample-efficient approach to optimizing ML-based traffic
analysis pipelines. \sysname surpasses 0.99~HVI (using a worst-case reference
point) within 87~iterations on average compared to \sysnamebase's
240~iterations, demonstrating a speedup of 2.76$\times$. This speedup can be
attributed to the incorporation of priors on the optimization parameters as
described in Section~\ref{sec:preprocessing}, which helps \sysname emphasize
more promising regions in the search space. \simanneal and \randsearch are less
sample efficient, surpassing 0.99~HVI at $1{,}295$~iterations and
$1{,}469$~iterations for a \sysname speedup of 14.9$\times$ and 16.9$\times$,
respectively.

\subsection{Ablation Study of the Profiler}
\label{sec:profiler_eval}

We assess the impact of the Profiler on the estimated Pareto front found by
\sysname. We retain the Optimizer, including dimensionality reduction and prior
injection, and perform an ablation study by replacing $\texttt{cost}(x)$ and
$\texttt{perf}(x)$ measurements with heuristic metrics.  We devise four variants
of \sysname. The first is \catonaivecost, which replaces the original cost
metric with the sum of the costs of each feature in isolation.  This design
captures the end-to-end systems costs of individual features, but fails to
account for shared processing steps during packet capture and feature
extraction. The second is \catomodelinfcost, which measures the model inference
speed but ignores the cost of packet capture and feature extraction.
\catopktdepthcost directly uses packet depth as the cost. \catonaiveperf retains
the original $\texttt{cost}(x)$ metric but replaces $\texttt{perf}(x)$ with the
sum of each feature's mutual information with respect to the target variable.
This version does not account for the effects of feature
interactions.\looseness=-1

\begin{figure}
    \centering
    \includegraphics[width=\columnwidth]{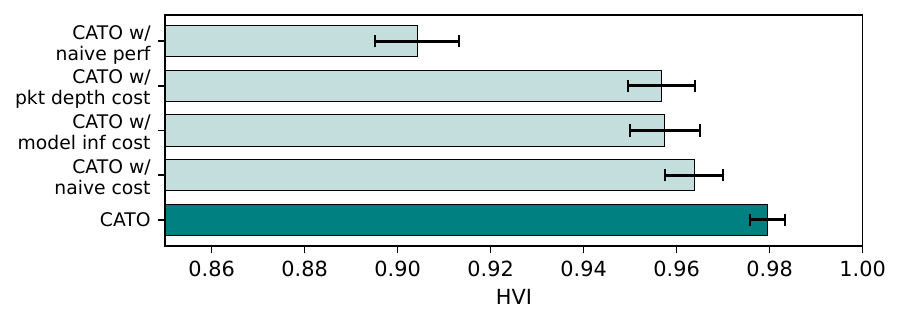}
	\vspace{-15pt}
    \caption{\sysname with alternative Profiler metrics. End-to-end measurements prove useful for estimating the Pareto front and validating the performance of identified solutions.}
    \label{fig:profiler_ablation}
\end{figure}

We run each variant for 50~Optimizer iterations using the smaller candidate
feature set, then measure (using the Profiler) the true $\texttt{perf}(x)$ and
$\texttt{cost}(x)$ of each sampled point in a post-processing step to compare
HVI. Figure~\ref{fig:profiler_ablation} reveals that \sysname comes closest to
the true Pareto front, demonstrating that there is value to incorporating real
model performance and systems costs measurements as feedback for the Optimizer.
Furthermore, we note that none of the variants provide a means to
\emph{validate} the expected in-network performance of their identified
solutions. In particular, \catonaiveperf and \catopktdepthcost do not yield
meaningful performance metrics. \catonaivecost and \catomodelinfcost are better,
but may overestimate or underestimate the true systems cost, respectively.
Depending on the concrete definition of $\texttt{cost}(x)$, this could lead to
the deployment of an unrealizable model (e.g., by overestimating the throughput
or underestimating the latency of the pipeline).\looseness=-1

\subsection{Microbenchmarks}
\label{sec:microbenchmarks}

In this section, we evaluate \sysname's robustness to varying search space sizes
and its sensitivity with respect to its BO initialization and damping
coefficient hyperparameters.

\myparab{Varying Maximum Connection Depth.}
\label{sec:vary_max_pkt_depth}
Previously, we showed that \sysname is highly effective across both large (67)
and small (6) candidate feature sets, but limited the search space to a
maximum connection depth of 50~packets. Here, we explore the impact of maximum
connection depth and the size of the search space on \sysname's ability to
identify Pareto-optimal feature representations.

In general, limiting the search space inherently restricts the number of
possible feature representations, potentially compromising the quality of the
estimated Pareto front. Conversely, expanding the search space offers a richer
set of feature representations, but makes locating the true Pareto front more
challenging. Table~\ref{tab:vary_max_depth} shows \sysname's performance across
different maximum packet depths for \iotclass using the full 67~candidate
features. Since it is difficult to compare HVI across configurations without a
ground truth, we report metrics for the estimated Pareto-optimal representations
with the highest F1 score and lowest overall execution time.

\begin{table}
    \centering
    \footnotesize
	\setlength{\tabcolsep}{3pt}

    \begin{tabular}{lrrrrrr}
    \toprule
    \multicolumn{1}{c|}{\textbf{Max}} & \multicolumn{3}{c|}{\textbf{Highest F1 Score}}                            & \multicolumn{3}{c}{\textbf{Lowest Execution Time}}          \\
    \multicolumn{1}{c|}{\textbf{Depth $N$}}   & $n$                    & F1 score             & \multicolumn{1}{r|}{Time ($\mu$s)} & $n$                    & F1 score             & Time ($\mu$s)               \\ \toprule
    \multicolumn{1}{r|}{3}                & 3                    & 0.959                & \multicolumn{1}{r|}{1.37}   & 1                    & 0.310                & 0.2                  \\
    \multicolumn{1}{r|}{5}                & 4                    & 0.983                & \multicolumn{1}{r|}{1.30}   & 1                    & 0.520                & 0.26                 \\
    \multicolumn{1}{r|}{10}               & 7                    & 0.994                & \multicolumn{1}{r|}{2.04}   & 1                    & 0.520                & 0.26                 \\
    \multicolumn{1}{r|}{25}               & 7                    & 0.989                & \multicolumn{1}{r|}{2.61}   & 1                    & 0.520                & 0.26                 \\
    \multicolumn{1}{r|}{50}               & 7                    & 0.993                & \multicolumn{1}{r|}{2.10}   & 1                    & 0.461                & 0.27                 \\
    \multicolumn{1}{r|}{100}              & 10                   & 0.990                & \multicolumn{1}{r|}{3.01}   & 1                    & 0.005                & 0.24                 \\
    \multicolumn{1}{r|}{$\infty$}         & 42k                  & 0.984                & \multicolumn{1}{r|}{28.2}   & 52k                  & 0.944                & 16.5                 \\ \bottomrule
                                        
    \end{tabular}

	\vspace{-5pt}
\caption{Estimated Pareto-optimal solutions with the highest F1 score and lowest execution time for different maximum packet depths. \sysname is able to identify high quality solutions, even when expanding the maximum connection depth.}
\label{tab:vary_max_depth}
\end{table}

We can see that restricting the packet depth to very small values (e.g., less
than 5) limits \sysname's ability to find a solution that yields an F1 score
above 0.99, likely because not many feature sets can achieve such high model
performance using only the first few packets in a flow. However, if we expand
the search space to include features extracted from \emph{up to} the first
10--100~packets, \sysname is still able to identify feature sets that only need
the first 7--10~packets to achieve good model performance despite the larger
maximum packet depth. However, if the search space over packet depth is
unbounded, \sysname struggles converge to a feature representation with low cost
since the Optimizer has too much flexibility to explore any value between 1 and
the maximum number of packets across \emph{all} flows in the training set.
These results reinforce that concurrently searching over different features {and
when to collect those features}, rather than predefining a fixed connection
depth, can lead to more optimal traffic analysis pipelines. \sysname can
identify highly efficient and predictive feature representations within a wide
range of connection depths, even without prior knowledge of its optimum.

\begin{figure}
    \centering
	\begin{subfigure}[t]{0.49\columnwidth}
        \includegraphics[width=\linewidth]{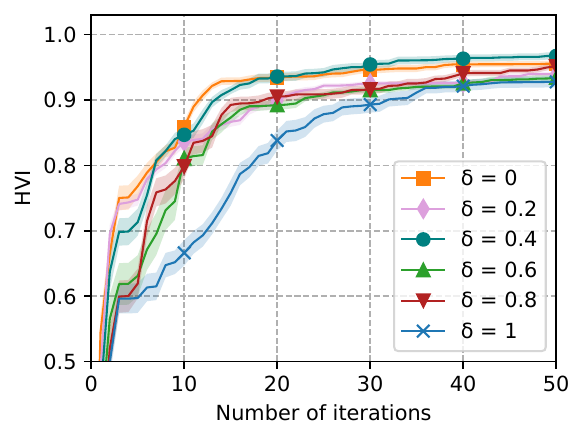}
        \caption{Damping coefficient $\delta$.}
        \label{fig:damping}
    \end{subfigure}
	\begin{subfigure}[t]{0.49\columnwidth}
        \includegraphics[width=\linewidth]{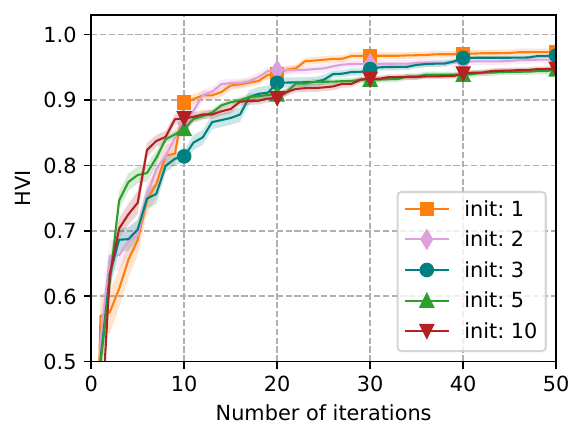}
        \caption{BO initialization samples.}
        \label{fig:init}
    \end{subfigure}
    \caption{Effects of varying the damping coefficient and the number of samples used to initialize the BO surrogate model.}
    \label{fig:sensitivity}
\end{figure}

\myparab{Sensitivity Analysis.}
We analyze the sensitivity of \sysname with respect to its hyperparameters. We
again use the smaller candidate feature set to measure HVI against the true
Pareto front with a worst-case reference point. Figure~\ref{fig:damping} shows
the impact of varying the damping coefficient $\delta$ between 0~and~1. Recall
that $\delta = 0$ represents a prior probability equivalent to the normalized
mutual information, while $\delta = 1$ represents uniform priors for each
feature. Using uniform priors performs the worst with an HVI of 0.93 after
50~iterations. With less damping, \sysname converges faster in earlier
iterations. We find that a $\delta=0.4$ results in the highest performance at
50~iterations, while $\delta = 0$ also performs well overall.

In Figure~\ref{fig:init}, we vary the number of samples used to initialize the
BO surrogate model. Initialization samples are chosen at random but weighted
according to the priors. We observe little difference in \sysname's performance
for small initialization values, but find that initializing with just 1~point
empirically results in the highest HVI after 50~iterations. We choose to go with
a more conventional value for BO~\cite{jones1998bo}, and choose 3~initialization
samples by default.

\section{Related Work}
\label{sec:related}

\myparab{Efficient Inference.}
Several systems have been proposed for efficient traffic analysis inference.
However, many of these efforts focus only on increasing the speed of the final
model inference stage, rather than that of the end-to-end serving
pipeline~\cite{liang2019neural,tong2014high,devprasad2022context}. Such
approaches overlook the effects of the packet capture and feature extraction
stages, both of which \sysname considers in its end-to-end optimization.
AC-DC~\cite{jiang2023acdc} and pForest~\cite{bussegrawitz2019pforest} are
similar in that they explicitly consider the preprocessing costs of extracting
features. AC-DC performs inference under dynamic memory constraints while
pForest targets programmable hardware. Both of these differ from \sysname in
that they generate pools of models and dynamically switch between them based on
inference requirements.

There is a growing body of research that proposes the use of programmable
hardware for traffic analysis~\cite{jafri2024leo, sanvito2018networkai,
kamath2015p4classification, xiong2019ilsy, siracusano2022n3ic, swamy2022taurus,
zhou2023netbeacon, yan2024bos, zheng2022planter,
swamy2023homunculus,barradas2021flowlens, bussegrawitz2019pforest}. For example,
N3IC~\cite{siracusano2022n3ic} uses binary neural networks to implement traffic
analysis models on FPGAs and SmartNICs, while BoS~\cite{yan2024bos} enables RNN
inference on programmable switches. These approaches focus on the trade-off
between accuracy and efficiency of the \emph{model} under the constraints of
dataplane hardware, such as limited memory and lack of support for
floating-point operations, multiplications, and loops. Our work, by contrast,
focuses on the choice of \emph{traffic} representation---spanning both feature
selection and connection depth---to co-optimize systems performance and
predictive accuracy. \sysname can complement hardware-focused techniques: by
applying the \sysname Optimizer and extending our Profiler implementation to
target programmable hardware, we can identify traffic representations that are
cheaper to collect while also validating the end-to-end performance of the
hardware pipeline. This combined strategy can further boost the efficiency and
predictive performance of ML-based traffic analysis, which we leave for future
work.

\myparab{Balancing Systems and Model Performance.} There exist general-purpose
serving frameworks that aim to balance efficiency with model performance through
techniques like caching~\cite{crankshaw2017clipper}, adaptive model
selection~\cite{crankshaw2017clipper, romero2021infaas},
autoscaling~\cite{crankshaw2020inferline, romero2021infaas,zhang2019mark}, and
scheduling for resource-quality
trade-offs~\cite{zhang2017videostorm,li2023alpaserve}.
Unlike \sysname, these frameworks primarily focus on optimizing resource
allocation and are less suited for real-time traffic analysis, where even small
inefficiencies in end-to-end systems performance can cause not just delayed
results, but also potentially invalidate the model due to packet loss.
Additionally, \sysname explicitly considers the optimal point in the flow (i.e.,
connection depth) for making predictions, which previous frameworks do not
address. Traffic Refinery~\cite{bronzino2021trafficrefinery} considers joint
optimization of model and system performance for ML-based network traffic
analysis. However, it relies on manual exploration of data representations and
focuses only on individual feature costs, whereas \sysname performs automated
end-to-end optimization of the entire analysis pipeline.

\myparab{Bayesian Optimization for Traffic Analysis.}
Bayesian Optimization is a popular technique for compiler
optimization~\cite{hellsten2023baco}, FPGA design~\cite{nardi2019hypermapper},
and hyperparameter tuning~\cite{snoek2012mlbo,turner2021hpobo}, but has seen
limited use in optimizing ML-based traffic analysis pipelines. Most similar is
Homunculus~\cite{swamy2023homunculus}, which uses Bayesian optimization to
generate ML models for datacenter network applications under the resource
constraints of data-plane hardware. Homunculus optimizes over different model
architectures and their hyperparameters, but is single-objective and does not
consider the effects of different feature sets, connection depths, and their
associated systems costs. \sysname, on the other hand, uses multi-objective BO
over the entire feature representation space to simultaneously optimize serving
efficiency and predictive performance.

\section{Conclusion}

In this work, we introduced \sysname, a framework for end-to-end optimization of
ML-based traffic analysis pipelines. By leveraging multi-objective Bayesian
optimization coupled with a realistic pipeline generator and profiler, \sysname
efficiently builds and validates serving pipelines that balance both model
accuracy and systems performance. Our evaluations on live traffic and offline
traces showed that \sysname can improve end-to-end inference latency,
throughput, and pipeline execution time across diverse traffic analysis tasks,
while also maintaining or enhancing model performance. Future work includes
broader model selection strategies and extending performance profiling across
heterogeneous serving hardware.

\section*{Acknowledgements} We thank our shepherd Zhizhen Zhong as well as Tina
Wu, Thea Rossman, Qizheng Zhang, Carl Hvarfner, Luigi Nardi, and the anonymous
reviewers for their helpful feedback. We thank the Stanford networking team,
including Andrej Krevl, Johan van Reijendam, and Will Johnson. This work was
supported in part by a Sloan Research Fellowship, the National Science
Foundation under Grant Numbers \#2319080 and \#2124424, the ANR Project No
ANR-21-CE94-0001-01 (MINT), and gifts from Google, Inc., Cisco Systems, Inc.,
and Comcast Corporation. Any opinions, findings, and conclusions or
recommendations expressed in this material are those of the author(s) and do not
necessarily reflect the views of the NSF or other funding
organizations.\looseness=-1

\label{endOfBody}

{
\footnotesize
\bibliographystyle{plain}
\bibliography{reference.bib}

\begin{thebibliography}{10}

\bibitem{smartcore}
Smartcore.
\newblock \url{https://smartcorelib.org/}, 2023.

\bibitem{azurewaf}
Web application firewall documentation.
\newblock
  \url{https://learn.microsoft.com/en-us/azure/web-application-firewall/},
  2023.

\bibitem{zeek}
Zeek.
\newblock \url{https://zeek.org/}, 2023.

\bibitem{abbasi2021dlsurvey}
Mahmoud Abbasi, Amin Shahraki, and Amir Taherkordi.
\newblock Deep learning for network traffic monitoring and analysis ({NTMA}): A
  survey.
\newblock In {\em Computer Communications}, 2021.

\bibitem{alan2016androidid}
Hasan~Faik Alan and Jasleen Kaur.
\newblock Can android applications be identified using only {TCP/IP} headers of
  their launch time traffic?
\newblock In {\em ACM Conference on Security and Privacy in Wireless Networks},
  2016.

\bibitem{altmann2010pimp}
Andr\'{e} Altmann, Laura Toloşi, Oliver Sander, and Thomas Lengauer.
\newblock Permutation importance: A corrected feature importance measure.
\newblock In {\em Bioinformatics}, 2010.

\bibitem{aouini2022nfstream}
Zied Aouini and Adrian Pekar.
\newblock {NFStream}: A flexible network data analysis framework.
\newblock In {\em Computer Networks}, 2022.

\bibitem{balachandran2013qoe}
Athula Balachandran, Vyas Sekar, Aditya Akella, Srinivasan Seshan, Ion Stoica,
  and Hui Zhang.
\newblock Developing a predictive model of quality of experience for internet
  video.
\newblock In {\em ACM Special Interest Group on Data Communication (SIGCOMM)},
  2013.

\bibitem{barradas2021flowlens}
Diogo Barradas, Nuno Santos, Lu\'{i}s Rodrigues, Salvatore Signorello, Fernando
  M.~V. Ramos, and Andr\'{e} Madeira.
\newblock Flowlens: Enabling efficient flow classification for ml-based network
  security applications.
\newblock In {\em Network and Distributed Systems Security Symposium (NDSS)},
  2021.

\bibitem{bernaille2006onthefly}
Laurent Bernaille, Renata Teixeira, Ismael Akodjenou, Augustin Soule, and
  Kav\'{e} Salamatian.
\newblock Traffic classification on the fly.
\newblock In {\em ACM SIGCOMM Computer Communication Review}, 2006.

\bibitem{bernaille2006earlyappid}
Laurent Bernaille, Renata Teixeira, and Kav\'{e} Salamatian.
\newblock Early application identification.
\newblock In {\em International Conference on Emerging Networking Experiments
  and Technologies (CoNEXT)}, 2006.

\bibitem{boutaba2018mlsurvey}
Raouf Boutaba, Mohammad~A. Salahuddin, Noura Limam, Sara Ayoubi, Nashid
  Shahriar, Felipe Estrada-Solano, and Oscar~M. Caicedo.
\newblock A comprehensive survey on machine learning for networking: evolution,
  applications and research opportunities.
\newblock In {\em Journal of Internet Services and Applications}, 2018.

\bibitem{bouthiller2020mlsurvey}
Xavier Bouthillier and Ga\"{e}l Varoquaux.
\newblock Survey of machine-learning experimental methods at {NeurIPS2019 and
  ICLR2020}.
\newblock {\em Research Report, Inria Saclay Ile de France}, 2020.

\bibitem{bronzino2021trafficrefinery}
Francesco Bronzino, Paul Schmitt, Sara Ayoubi, Hyojoon Kim, Renata Teixeira,
  and Nick Feamster.
\newblock Traffic refinery: Cost-aware data representation for machine learning
  on network traffic.
\newblock In {\em ACM Measurement and Analysis of Computing Systems}, 2021.

\bibitem{bronzino2019inferring}
Francesco Bronzino, Paul Schmitt, Sara Ayoubi, Guilherme Martins, Renata
  Teixeira, and Nick Feamster.
\newblock Inferring streaming video quality from encrypted traffic: Practical
  models and deployment experience.
\newblock In {\em ACM Measurement and Analysis of Computing Systems}, 2019.

\bibitem{bussegrawitz2019pforest}
Coralie Busse-Grawitz, Roland Meier, Alexander Dietm\"{u}ller, Tobias
  B\"{u}hler, and Laurent Vanbever.
\newblock pforest: In-network inference with random forests.
\newblock {\em arXiv preprint arXiv:1909.05680v2}, 2022.

\bibitem{calandra2014robotbo}
Roberto Calandra, Nakul Gopalan, Andr\'{e} Seyfarth, Jan Peters, and Marc~Peter
  Deisenroth.
\newblock {B}ayesian gait optimization for bipedal locomotion.
\newblock In {\em International Conference on Learning and Intelligent
  Optimization (LION)}, 2014.

\bibitem{chang2023llb}
Briang Chang, Kausik Subramanian, Loris D'Antoni, and Aditya Akella.
\newblock Learned load balancing.
\newblock In {\em International Conference on Distributed Computing and
  Networking (ICDCN)}, 2023.

\bibitem{crankshaw2020inferline}
Daniel Crankshaw, Gur-Eyal Sela, Xiangxi Mo, Corey Zumar, Ion Stoica, Joseph
  Gonzalez, and Alexey Tumanov.
\newblock Inferline: Latency-aware provisioning and scaling for prediction
  serving pipelines.
\newblock In {\em ACM Symposium on Cloud Computing}, 2020.

\bibitem{crankshaw2017clipper}
Daniel Crankshaw, Xin Wang, Giulio Zhou, Michael~J. Franklin, Joseph~E.
  Gonzalez, and Ion Stoica.
\newblock Clipper: A low-latency online prediction serving system.
\newblock In {\em USENIX Symposium on Networked Systems Design and
  Implementation (NSDI)}, 2017.

\bibitem{dainotti2015earlymulticlass}
Alberto Dainotti, Antonio Pescap\'{e}, and Carlo Sansone.
\newblock Early classification of network traffic through multi-classification.
\newblock In {\em International Workshop on Traffic Monitoring and Analysis
  (TMA)}, 2011.

\bibitem{devprasad2022context}
Kayathri~Devi Devprasad, Sukumar Ramanujam, and Suresh~Babu Rajendran.
\newblock Context adaptive ensemble classification mechanism with
  multi-criteria decision making for network intrusion detection.
\newblock {\em Concurrency and Computation: Practice and Experience}, 2022.

\bibitem{drapergil2016vpndetection}
Gerard Draper-Gil, Arash~Habibi Lashkari, Mohammad Saiful~Islam Mamun, and
  Ali~A. Ghorbani.
\newblock Characterization of encrypted and {VPN} traffic using time-related
  features.
\newblock In {\em International Conference on Information Systems Security and
  Privacy (ICISSP)}, 2016.

\bibitem{frazier2018botutorial}
Peter~I. Frazier.
\newblock A tutorial on {B}ayesian optimization.
\newblock {\em arXiv preprint arXiv:1807.02811}, 2018.

\bibitem{fu2021whisper}
Chuanpu Fu, Qi~Li, Meng Shen, and Ke~Xu.
\newblock Realtime robust malicious traffic detection via frequency domain
  analysis.
\newblock In {\em ACM SIGSAC Conference on Computer and Communication Security
  (CCS)}, 2021.

\bibitem{gutterman2020requet}
Craig Gutterman, Katherine Guo, Sarthak Arora, Xiaoyang Wang, Les Wu, Ethan
  Katz-Bassett, and Gil Zussman.
\newblock Requet: Real-time {QoE} detection for encrypted youtube traffic.
\newblock In {\em ACM Transactions on Multimedia Computing, Communications},
  2020.

\bibitem{guyon2002rfe}
Isabelle Guyon, Jason Weston, Stephen Barnhill, and Vladimir Vapnik.
\newblock Gene selection for cancer classification using support vector
  machines.
\newblock In {\em Machine Learning}, 2002.

\bibitem{hellsten2023baco}
Erik Hellsten, Artur Souza, Johannes Lenfers, Rubens Lacouture, Olivia Hsu,
  Adel Ejjeh, Fredrik Kjolstad, Michel Steuwer, Kunle Olukotun, and Luigi
  Nardi.
\newblock {BaCO}: A fast and portable {B}ayesian compiler optimization
  framework.
\newblock In {\em ACM Architectural Support for Programming Languages and
  Operating Systems}, 2023.

\bibitem{holland2021nprintml}
Jordan Holland, Paul Schmitt, Nick Feamster, and Prateek Mittal.
\newblock New directions in automated traffic analysis.
\newblock In {\em ACM Conference on Computer and Communication Security (CCS)},
  2021.

\bibitem{huang2013appr}
Nen-Fu Huang, Gin-Yuan Jai, Han-Chieh Chao, Yih-Jou Tzang, and Hong-Yi Chang.
\newblock Application traffic classification at the early stage by
  characterizing application rounds.
\newblock In {\em Information Sciences}, 2013.

\bibitem{hugon2023towards}
Johann Hugon, Gaetan Nodet, Anthony Busson, and Francesco Bronzino.
\newblock Towards adaptive ml traffic processing systems.
\newblock In {\em Proceedings of the on CoNEXT Student Workshop 2023}, 2023.

\bibitem{hvarfner2022pibo}
Carl Hvarfner, Danny Stoll, Artur Souza, Marius Lindauer, Frank Hutter, and
  Luigi Nardi.
\newblock {$\pi$BO}: Augmenting acquisition functions with user beliefs for
  {B}ayesian optimization.
\newblock In {\em International Conference on Learning Representations (ICLR)},
  2022.

\bibitem{jafri2024leo}
Syed~Usman Jafri, Sanjay Rao, Vishal Shrivastav, and Mohit Tawarmalani.
\newblock Leo: Online ml-based traffic classification at multi-terabit line
  rate.
\newblock In {\em USENIX Symposium on Networked Systems Design and
  Implementation (NSDI)}, 2024.

\bibitem{jiang2023acdc}
Xi~Jiang, Shinan Liu, Saloua Naama, Francesco Bronzino, Paul Schmitt, and Nick
  Feamster.
\newblock {AC-DC}: Adaptive ensemble classification for network traffic
  identification.
\newblock {\em arXiv preprint arXiv:2302.11718}, 2023.

\bibitem{jones1998bo}
Donald~R. Jones, Matthias Schonlau, and William~J. Welch.
\newblock Efficient global optimization of expensive black-box functions.
\newblock In {\em Journal of Global Optimization}, 1998.

\bibitem{kamath2015p4classification}
Radhakrishna Kamath and Krishna~M. Sivalingam.
\newblock Machine learning based flow classification in {DCN}s using {P4}
  switches.
\newblock In {\em International Conference on Computer Communications and
  Networks}, 2015.

\bibitem{karagiannis2005blinc}
Thomas Karagiannis, Konstantina Papagiannaki, and Michalis Faloutsos.
\newblock {BLINC}: Multilevel traffic classification in the dark.
\newblock In {\em ACM Special Interest Group on Data Communication (SIGCOMM)},
  2005.

\bibitem{kirkpatrick1983anneal}
Scott Kirkpatrick, C.~Daniel~Gelatt Jr., and Mario~P. Vecchi.
\newblock Optimization by simulated annealing.
\newblock In {\em Science}, 1983.

\bibitem{krishnamoorthi2017buffest}
Vengatanathan Krishnamoorthi, Niklas Carlsson, Emir Halepovic, and Eric
  Petajan.
\newblock {BUFFEST}: Predicting buffer conditions and real-time requirements of
  {HTTP(S)} adaptive streaming clients.
\newblock In {\em ACM Multimedia Systems Conference}, 2017.

\bibitem{lee2020switchtree}
Jong-Hyouk Lee and Kamal Singh.
\newblock {SwitchTree}: In-network computing and traffic analyses with random
  forests.
\newblock In {\em Neural Computing and Applications}, 2020.

\bibitem{li2019qualityeval}
Miqing Li and Xin Yao.
\newblock Quality evaluation of solution sets in multiobjective optimisation: A
  survey.
\newblock In {\em ACM Computing Surveys}, 2019.

\bibitem{li2023alpaserve}
Zhuohan Li, Lianmin Zheng, Yinmin Zhong, Vincent Liu, Ying Sheng, Xin Jin,
  Yanping Huang, Zhifeng Chen, Hao Zhang, Joseph~E Gonzalez, et~al.
\newblock Alpaserve: Statistical multiplexing with model parallelism for deep
  learning serving.
\newblock {\em USENIX Symposium on Operating Systems Design and
  Implementation}, 2023.

\bibitem{liang2019neural}
Eric Liang, Hang Zhu, Xin Jin, and Ion Stoica.
\newblock Neural packet classification.
\newblock In {\em SIGCOMM}, 2019.

\bibitem{liu2007kmeansclassification}
Yingqiu Liu, Wei Li, and Yunchun Li.
\newblock Network traffic classification using {K}-means clustering.
\newblock In {\em International Multi-Symposiums on Computer and Computational
  Sciences}, 2007.

\bibitem{lopezmartin2017nniot}
Manuel Lopez-Martin, Belen Carro, Antonio Sanchez-Esguevillas, and Jaime
  Lloret.
\newblock Network traffic classifier with convolutional and recurrent neural
  networks for internet of things.
\newblock In {\em IEEE Access}, 2017.

\bibitem{lotfollahi2020deep}
Mohammad Lotfollahi, Mahdi Jafari~Siavoshani, Ramin Shirali Hossein~Zade, and
  Mohammdsadegh Saberian.
\newblock Deep packet: A novel approach for encrypted traffic classification
  using deep learning.
\newblock {\em Soft Computing}, 2020.

\bibitem{mangla2019emimic}
Tarun Mangla, Emir Halepovic, Mostafa Ammar, and Ellen Zegura.
\newblock Using session modeling to estimate {HTTP}-based video {QoE} metrics
  from encrypted network traffic.
\newblock In {\em IEEE Transactions on Network and Service Management}, 2019.

\bibitem{mazhar2018qoe}
M.~Hammad Mazhar and Zubair Shafiq.
\newblock Real-time video quality of experience monitoring for {HTTPS} and
  {QUIC}.
\newblock In {\em IEEE Conference on Computer Communications (INFOCOM)}, 2018.

\bibitem{miettinen2017iotsentinel}
Markus Miettinen, Samuel Marchal, Ibbad Hafeez, Ahmad-Reza Sadeghi, N.~Asokan,
  and Sasu Tarkoma.
\newblock {IoT} sentinel: Automated device-type identification for security
  enforcement in {IoT}.
\newblock In {\em International Conference on Distributed Computing Systems},
  2017.

\bibitem{moore2005discriminators}
Andrew Moore, Denis Zuev, and Michael Crogan.
\newblock Discriminators for use in flow-based classification.
\newblock Technical report, 2005.

\bibitem{nardi2019hypermapper}
Luigi Nardi, Artur Souza, David Koeplinger, and Kunle Olukotun.
\newblock {HyperMapper}: a practical design space exploration framework.
\newblock In {\em IEEE International Symposium on Modeling, Analysis, and
  Simulation of Computer and Telecommunication Systems (MASCOTS)}, 2019.

\bibitem{nguyen2008mlsurvey}
Thuy~T.T. Nguyen and Grenville Armitage.
\newblock A comprehensive survey on machine learning for networking: Evolution,
  applications and research opportunities.
\newblock In {\em IEEE Communications Surveys \& Tutorials}, 2008.

\bibitem{peng2015packetnumber}
Lizhi Peng, Bo~Yang, and Yuehui Chen.
\newblock Effective packet number for early stage internet traffic
  identification.
\newblock In {\em Neurocomputing}, 2015.

\bibitem{piet2023ggfast}
Julien Piet, Dubem Nwoji, and Vern Paxson.
\newblock {GGFAST}: Automating generation of flexible network traffic
  classifiers.
\newblock In {\em ACM Special Interest Group on Data Communication (SIGCOMM)},
  2023.

\bibitem{rezaei2019deepappid}
Shahbaz Rezaei, Bryce Kroencke, and Xin Liu.
\newblock Large-scale mobile app identification using deep learning.
\newblock In {\em IEEE Access}, 2019.

\bibitem{rimmer2017automated}
Vera Rimmer, Davy Preuveneers, Marc Juarez, Tom Van~Goethem, and Wouter Joosen.
\newblock Automated website fingerprinting through deep learning.
\newblock {\em arXiv preprint arXiv:1708.06376}, 2017.

\bibitem{romero2021infaas}
Francisco Romero, Qian Li, Neeraja~J Yadwadkar, and Christos Kozyrakis.
\newblock {INFaaS}: Automated model-less inference serving.
\newblock In {\em USENIX Annual Technical Conference (USENIX ATC)}, 2021.

\bibitem{sanvito2018networkai}
Davide Sanvito, Giuseppe Siracusano, and Roberto Bifulco.
\newblock Can the network be the ai accelerator?
\newblock In {\em Morning Workshop on In-Network Computing}, 2018.

\bibitem{sena2009earlysvm}
Gabriel~G\'{o}mez Sena and Pablo Belzarena.
\newblock Early traffic classification using support vector machines.
\newblock In {\em International Latin American Networking Conference (LANC)},
  2009.

\bibitem{shahriari2016boreview}
Bobak Shahriari, Kevin Swersky, Ziyu Wang, Ryan~P. Adams, and Nando de~Freitas.
\newblock Taking the human out of the loop: A review of bayesian optimization.
\newblock In {\em Proceedings of IEEE, vol. 104, no. 1}, 2016.

\bibitem{shapira2021flowpic}
Tal Shapira and Yuval Shavitt.
\newblock {FlowPic}: A generic representation for encrypted traffic
  classification and applications identification.
\newblock In {\em IEEE Transactions on Network and Service Management}, 2021.

\bibitem{singh2013mlsurvey}
Jayveer Singh and Manisha Nene.
\newblock A survey on machine learning techniques for intrusion detection
  systems.
\newblock In {\em Intl.\ Journal of Advanced Research in Computer and
  Communication Engineering}, 2013.

\bibitem{siracusano2022n3ic}
Giuseppe Siracusano, Salvator Galea, Davide Sanvito, Mohammad Malekzadeh,
  Gianni Antichi, Paolo Costa, Hamed Haddadi, and Roberto Bifulco.
\newblock Re-architecting traffic analysis with neural network interface cards.
\newblock In {\em USENIX Symposium on Networked Systems Design and
  Implementation (NSDI)}, 2022.

\bibitem{sivanathan2019iotclass}
Arunan Sivanathan, Hassan~Habibi Gharakheili, Franco Loi, Adam Radford, Chamith
  Wijenayake, Arun Vishwanath, and Vijay Sivaraman.
\newblock Classifying {IoT} devices in smart environments using network traffic
  characteristics.
\newblock In {\em IEEE Transactions on Mobile Computing}, 2019.

\bibitem{snoek2012mlbo}
Jasper Snoek, Hugo Larochelle, and Ryan~P. Adams.
\newblock Practical {B}ayesian optimization of machine learning algorithms.
\newblock In {\em International Conference on Advances in Neural Information
  Processing Systems}, 2012.

\bibitem{swamy2022taurus}
Tushar Swamy, Alexander Rucker, Muhammad Shahbaz, Ishan Gaur, and Kunle
  Olukotun.
\newblock Taurus: A data plane architecture for per-packet {ML}.
\newblock In {\em ACM Architectural Support for Programming Languages and
  Operating Systems}, 2022.

\bibitem{swamy2023homunculus}
Tushar Swamy, Annus Zulfiqar, Luigi Nardi, Muhammad Shahbaz, and Kunle
  Olukotun.
\newblock Homunculus: Auto-generating efficient data-plane {ML} pipelines for
  datacenter networks.
\newblock In {\em ACM Architectural Support for Programming Languages and
  Operating Systems}, 2023.

\bibitem{tong2014high}
Da~Tong, Yun~R Qu, and Viktor~K Prasanna.
\newblock High-throughput traffic classification on multi-core processors.
\newblock In {\em IEEE International Conference on High Performance Switching
  and Routing}, 2014.

\bibitem{turner2021hpobo}
Ryan Turner, David Eriksson, Michael McCourt, Juha Kiili, Eero Laaksonen, Zhen
  Xu, and Isabelle Guyon.
\newblock Bayesian optimization is superior to random search for machine
  learning hyperparameter tuning: Analysis of the black-box optimization
  challenge 2020.
\newblock In {\em Proceedings of the NeurIPS 2020 Competition and Demonstration
  Track,}, 2021.

\bibitem{vergara2014mi}
Jorge~R. Vergara and Pablo~A. Est\'{e}vez.
\newblock A review of feature selection methods based on mutual information.
\newblock In {\em Neural Computing and Applications}, 2014.

\bibitem{wan2022retina}
Gerry Wan, Fengchen Gong, Tom Barbette, and Zakir Durumeric.
\newblock Retina: Analyzing 100 {GbE} traffic on commodity hardware.
\newblock In {\em ACM Special Interest Group on Data Communication (SIGCOMM)},
  2022.

\bibitem{wang2017malwareclassification}
Wei Wang, Ming Zhu, Xuewen Zeng, Xiaozhou Ye, and Yiqiang Sheng.
\newblock Malware traffic classification using convolutional neural network for
  representation learning.
\newblock In {\em International Conference on Information Networking}, 2017.

\bibitem{xiong2019ilsy}
Zhaoqi Xiong and Noa Zilberman.
\newblock Do switches dream of machine learning?: Toward in-network
  classification.
\newblock In {\em ACM Workshop on Hot Topics in Networks}, 2019.

\bibitem{yan2024bos}
Jinzhu Yan, Haotian Xu, Zhuotao Liu, Qi~Li, Ke~Xu, Mingwei Xu, and Jianping Wu.
\newblock Brain-on-switch: Towards advanced intelligent network data plane via
  nn-driven traffic analysis at line-speed.
\newblock In {\em USENIX Symposium on Networked Systems Design and
  Implementation}, 2024.

\bibitem{yang2021tfidf}
Hao Yang, Qin He, Zhenyan Liu, and Qian Zhang.
\newblock Malicious encryption traffic detection based on {NLP}.
\newblock In {\em Security and Communication Networks}, 2021.

\bibitem{yang2021noveltydetection}
Kun Yang, Nick Feamster, and Samory Kpotufe.
\newblock Feature extraction for novelty detection in network traffic.
\newblock {\em arXiv preprint arXiv:2006.16993v2}, 2021.

\bibitem{zhang2019mark}
Chengliang Zhang, Minchen Yu, Wei Wang, and Feng Yan.
\newblock {MArk}: Exploiting cloud services for cost-effective, {SLO}-aware
  machine learning inference serving.
\newblock In {\em USENIX Annual Technical Conference}, 2019.

\bibitem{zhang2017videostorm}
Haoyu Zhang, Ganesh Ananthanarayanan, Peter Bodik, Matthai Philipose, Paramvir
  Bahl, and Michael~J. Freedman.
\newblock Live video analytics at scale with approximation and delay-tolerance.
\newblock In {\em USENIX Symposium on Networked Systems Design and
  Implementation}, 2017.

\bibitem{zheng2022planter}
Changgang Zheng, Mingyuan Zang, Xinpeng Hong, Riyad Bensoussane, Shay
  Vargaftik, Yaniv Ben-Itzhak, and Noa Zilberman.
\newblock Automating in-network machine learning.
\newblock {\em arXiv preprint arXiv:2205.08824v1}, 2022.

\bibitem{zheng2022mtt}
Weiping Zheng, Jianhao Zhong, Qizhi Zhang, and Gansen Zhao.
\newblock {MTT}: An efficient model for encrypted network traffic
  classification using multi-task transformer.
\newblock {\em Applied Intelligence}, 2022.

\bibitem{zhou2023netbeacon}
Guangmeng Zhou, Zhuotao Liu, Chuanpu Fu, Qi~Li, and Ke~Xu.
\newblock An efficient design of intelligent network data plane.
\newblock In {\em USENIX Security Symposium}, 2023.

\end{thebibliography}
}

\appendix

\section{Candidate Features}
\label{app:candidate_features}

Table~\ref{tab:candidate_features} lists the set of candidate features used in
our evaluations. We aggregate common network flow features used throughout
networking research for ML-based traffic analysis~\cite{siracusano2022n3ic,
bussegrawitz2019pforest, wan2022retina, swamy2023homunculus,
moore2005discriminators, zeek, bronzino2019inferring}.  These are also commonly
supported by open source tools, and are not specific to any particular use case.
However, for privacy reasons related to our live traffic experiments, we
restrict these to various types of summary statistics to avoid saving packet
payloads to disk.

\begin{table}
\scriptsize

    \begin{tabular}{llr}
    \textbf{Feature}          & \textbf{Description}                        & \textbf{\begin{tabular}[c]{@{}r@{}}In mini \\ cand. set\end{tabular}} \\ \toprule
    \texttt{dur}              & total duration                              & yes                                                                     \\
    \texttt{proto}            & transport layer protocol                    & no                                                                     \\
    \texttt{s\_port}          & src port                                    & no                                                                     \\
    \texttt{d\_port}          & dst port                                    & no                                                                     \\
    \texttt{s\_load}          & src $\rightarrow$ dst bps                               & yes                                                                     \\
    \texttt{d\_load}          & dst $\rightarrow$ src bps                               & no                                                                     \\
    \texttt{s\_pkt\_cnt}      & src $\rightarrow$ dst packet count                      & yes                                                                     \\
    \texttt{d\_pkt\_cnt}      & dst $\rightarrow$ src packet count                      & no                                                                     \\
    \texttt{tcp\_rtt}         & time between SYN and ACK                    & no                                                                     \\
    \texttt{syn\_ack}         & time between SYN and SYN/ACK                & no                                                                     \\
    \texttt{ack\_dat}         & time between SYN/ACK and ACK                & no                                                                     \\
    \texttt{s\_bytes\_sum}    & src $\rightarrow$ dst total bytes                       & yes                                                                     \\
    \texttt{d\_bytes\_sum}    & dst $\rightarrow$ src total bytes                       & no                                                                     \\
    \texttt{s\_bytes\_mean}   & src $\rightarrow$ dst mean packet size                  & yes                                                                     \\
    \texttt{d\_bytes\_mean}   & dst $\rightarrow$ src mean packet size                  & no                                                                     \\
    \texttt{s\_bytes\_min}    & src $\rightarrow$ dst min packet size                   & no                                                                     \\
    \texttt{d\_bytes\_min}    & dst $\rightarrow$ src min packet size                   & no                                                                     \\
    \texttt{s\_bytes\_max}    & src $\rightarrow$ dst max packet size                   & no                                                                     \\
    \texttt{d\_bytes\_max}    & dst $\rightarrow$ src max packet size                   & no                                                                     \\
    \texttt{s\_bytes\_med}    & src $\rightarrow$ dst median packet size                & no                                                                     \\
    \texttt{d\_bytes\_med}    & dst $\rightarrow$ src median packet size                & no                                                                     \\
    \texttt{s\_bytes\_std}    & src $\rightarrow$ dst std dev packet size               & no                                                                     \\
    \texttt{d\_bytes\_std}    & dst $\rightarrow$ src std dev packet size               & no                                                                     \\
    \texttt{s\_iat\_sum}      & src $\rightarrow$ dst total packet inter-arrival time   & no                                                                     \\
    \texttt{d\_iat\_sum}      & dst $\rightarrow$ src total packet inter-arrival time   & no                                                                     \\
    \texttt{s\_iat\_mean}     & src $\rightarrow$ dst mean packet inter-arrival time    & yes                                                                     \\
    \texttt{d\_iat\_mean}     & dst $\rightarrow$ src mean packet inter-arrival time    & no                                                                     \\
    \texttt{s\_iat\_min}      & src $\rightarrow$ dst min packet inter-arrival time     & no                                                                     \\
    \texttt{d\_iat\_min}      & dst $\rightarrow$ src min packet inter-arrival time     & no                                                                     \\
    \texttt{s\_iat\_max}      & src $\rightarrow$ dst max packet inter-arrival time     & no                                                                     \\
    \texttt{d\_iat\_max}      & dst $\rightarrow$ src max packet inter-arrival time     & no                                                                     \\
    \texttt{s\_iat\_med}      & src $\rightarrow$ dst median packet inter-arrival time  & no                                                                     \\
    \texttt{d\_iat\_med}      & dst $\rightarrow$ src median packet inter-arrival time  & no                                                                     \\
    \texttt{s\_iat\_std}      & src $\rightarrow$ dst std dev packet inter-arrival time & no                                                                     \\
    \texttt{d\_iat\_std}      & dst $\rightarrow$ src std dev packet inter-arrival time & no                                                                     \\
    \texttt{s\_winsize\_sum}  & src $\rightarrow$ dst sum of TCP window sizes           & no                                                                     \\
    \texttt{d\_winsize\_sum}  & dst $\rightarrow$ src sum of TCP window sizes           & no                                                                     \\
    \texttt{s\_winsize\_mean} & src $\rightarrow$ dst mean TCP window size              & no                                                                     \\
    \texttt{d\_winsize\_mean} & dst $\rightarrow$ src mean TCP window size              & no                                                                     \\
    \texttt{s\_winsize\_min}  & src $\rightarrow$ dst min TCP window size               & no                                                                     \\
    \texttt{d\_winsize\_min}  & dst $\rightarrow$ src min TCP window size               & no                                                                     \\
    \texttt{s\_winsize\_max}  & src $\rightarrow$ dst max TCP window size               & no                                                                     \\
    \texttt{d\_winsize\_max}  & dst $\rightarrow$ src max TCP window size               & no                                                                     \\
    \texttt{s\_winsize\_med}  & src $\rightarrow$ dst med TCP window size               & no                                                                     \\
    \texttt{d\_winsize\_med}  & dst $\rightarrow$ src med TCP window size               & no                                                                     \\
    \texttt{s\_winsize\_std}  & src $\rightarrow$ dst std dev TCP window size           & no                                                                     \\
    \texttt{d\_winsize\_std}  & dst $\rightarrow$ src std dev TCP window size           & no                                                                     \\
    \texttt{s\_ttl\_sum}      & src $\rightarrow$ dst sum of IP TTL values               & no                                                                     \\
    \texttt{d\_ttl\_sum}      & dst $\rightarrow$ dst sum of IP TTL values               & no                                                                     \\
    \texttt{s\_ttl\_mean}     & src $\rightarrow$ dst mean TTL                          & no                                                                     \\
    \texttt{d\_ttl\_mean}     & dst $\rightarrow$ src mean TTL                          & no                                                                     \\
    \texttt{s\_ttl\_min}      & src $\rightarrow$ dst min TTL                           & no                                                                     \\
    \texttt{d\_ttl\_min}      & dst $\rightarrow$ src min TTL                           & no                                                                     \\
    \texttt{s\_ttl\_max}      & src $\rightarrow$ dst max TTL                           & no                                                                     \\
    \texttt{d\_ttl\_max}      & dst $\rightarrow$ src max TTL                           & no                                                                     \\
    \texttt{s\_ttl\_med}      & src $\rightarrow$ dst median TTL                        & no                                                                     \\
    \texttt{d\_ttl\_med}      & dst $\rightarrow$ src median TTL                        & no                                                                     \\
    \texttt{s\_ttl\_std}      & src $\rightarrow$ dst std dev TTL                       & no                                                                     \\
    \texttt{d\_ttl\_std}      & dst $\rightarrow$ src std dev TTL                           & no                                                                     \\
    \texttt{cwr\_cnt}         & number of packets with CWR flag set         & no                                                                     \\
    \texttt{ece\_cnt}         & number of packets with ECE flag set         & no                                                                     \\
    \texttt{urg\_cnt}         & number of packets with URG flag set         & no                                                                     \\
    \texttt{ack\_cnt}         & number of packets with ACK flag set         & no                                                                     \\
    \texttt{psh\_cnt}         & number of packets with PSH flag set         & no                                                                     \\
    \texttt{rst\_cnt}         & number of packets with RST flag set         & no                                                                     \\
    \texttt{syn\_cnt}         & number of packets with SYN flag set         & no                                                                     \\
    \texttt{fin\_cnt}         & number of packets with FIN flag set         & no                                                                     \\ \bottomrule
    \end{tabular}

\caption{Candidate feature set $\mathcal{F}$ containing 67~commonly used flow
features. We indicate the six that are used in the smaller candidate set for
ground truth analyses. }
\label{tab:candidate_features}
\end{table}

\section{Dataset Collection}
\label{app:datacollection_details}

All experiments with live traffic are performed on a server running
Ubuntu~20.04, with a dual Xeon Gold 6248R 3GHz CPU, 384~GB of memory, and a
100GbE Mellanox ConnectX-5 NIC. All experiments using offline datasets are
performed on a server with a dual Xeon Gold~6154 3GHz CPU and 384~GB of memory,
running Ubuntu~20.04.

\myparab{Web Application Classification.}
For \appclass, we classify the following applications: Netflix, Twitch,
Zoom, Microsoft Teams, Facebook, Twitter, or ``other.'' Connections are labeled
using the SNI from the TLS handshake, and only statistical features listed in
Table~\ref{tab:candidate_features} are collected (see Appendix~\ref{app:ethics} for ethical considerations involving live traffic). We use
Retina~\cite{wan2022retina} for data collection, which we run for 30~seconds
with 12.5\% flow sampling, and an additional 10~minutes with 50\% flow sampling
filtered on the target applications to help collect a more balanced dataset.
Flow sampling reduces the effective ingress network throughput while maintaining
per-connection consistency, and is done entirely in the NIC using hardware
filters~\cite{wan2022retina}. We avoid collecting at full network throughput to
ensure that no packets are dropped in the data collection phase. In total, we
collected 2M~samples of connection data over 50~different packet depths.

\myparab{IoT Device Recognition.}
For \iotclass, we classify one of 28~IoT device types using the UNSW
IoT dataset~\cite{sivanathan2019iotclass}. We use the September 2016 traces in
our evaluations, which include approximately 134K~connections. Models are trained and
evaluated using data from eight days of packet traces.

\myparab{Video Startup Delay Inference.}
We use the dataset collected by Bronzino et al.~\cite{bronzino2019inferring} for
\startupinf, focusing exclusively on YouTube traffic, where each video session
consists of a single TCP connection. The final dataset comprises
4,287~connections, capturing a wide range of startup delay times. Delay times
range from 315~ms to 54~seconds at P99, with the maximum observed delay being
14~minutes.

\section{Model Training}
\label{app:model_details}
For DT and RF model training, we use scikit-learn's DecisionTreeClassifier and
RandomForestClassifier with default parameters and tune the maximum tree depth
from the set \{3, 5, 10, 15, 20\}. The RandomForestClassifier is configured with
100~estimators. To integrate with the Rust-based packet capture and feature
extraction stages, the tuned DT and RF models are retrained in Rust using the
SmartCore~\cite{smartcore} library.

DNN hyperparameters are tuned over the following values: batch size \{16, 32,
64\}, learning rate \{0.001, 0.01\}, dropout rate \{0.2, 0.4, 0.6, 0.8\}, L2
regularization \{0.1, 0.5\}, and number of neurons in each hidden layer \{4, 8,
16\}.

\section{Measurement Details}
\label{app:measurement_details}

\myparab{End-to-End Inference Latency.}
We implement two versions of inference latency measurement, depending on the use
case. For \texttt{app-class}, we record the arrival timestamp of the SYN packet,
and subtract it from the timestamp of the final prediction output by the model.
Since we do not have access to live traffic from the \texttt{iot-class} dataset,
we compute the end-to-end inference latency by taking the sum of the pipeline
execution time, model inference time, and packet inter-arrival times up to the
specified connection depth. The inter-arrival times are calculated from packet
timestamps in the traces.

\myparab{Zero-Loss Throughput.}
Since we are unable to control the input traffic rate on our live network, we
choose to restrict our serving pipelines to a single core to differentiate the
performance of each optimization method. This ensures that we can find an upper
bound on the throughput since no solution will be able to saturate the input
traffic rate. To measure the zero-loss throughput, we leverage Retina's flow
sampling capabilities by starting at the full traffic rate and slowly decreasing
the percentage of flows randomly dropped by the NIC until we observe zero packet
loss for 30~seconds. We repeat this for multiple trials and take the average
zero-loss throughput sustained by the traffic analysis pipeline.

\myparab{Execution Time.}
Pipeline execution time is a measure of total CPU time spent in the serving
pipeline, excluding time spent waiting for packets to arrive. We measure this by
inserting calls to query the Read Time-Stamp Counter register at the start and
end of each packet processing step and the model inference stage.

\section{Optimization Wall-Clock Time} \label{app:wallclock} Wall-clock time
depends heavily on the application use-case, model type, number of samples
explored, and the concrete definitions of $\texttt{cost}(x)$ and
$\texttt{perf}(x)$. For reference, Table~\ref{tab:wallclock} reports the
breakdown in time elapsed for \sysname to compute the Pareto fronts depicted in
Figure~\ref{fig:compare_throughput_app} and Figure~\ref{fig:pareto_quality},
which target different use cases and system cost metrics. As expected, execution
time is mostly consumed by the Profiler, which, on each iteration, generates a
fresh serving pipeline, trains and evaluates the model, and measures the
end-to-end systems costs. We deem this a worthwhile trade-off because the time
spent by the Profiler to validate the systems cost of each sampled solution
ensures that the resulting Pareto-optimal pipeline can meet real-time
performance requirements. The BO-guided sampling that determines the next
feature representation for the Profiler to evaluate adds between 1.4 and
55~seconds per iteration depending on the search space.

\section{Reproducing Traffic Refinery}
\label{app:traffic_refinery}
We reproduce key components of the Traffic
Refinery~\cite{bronzino2021trafficrefinery} framework for evaluation. Traffic
Refinery defines several feature classes that contain features commonly used in
ML-based traffic analysis. Specifically, we replicate the PacketCounter
(\textsc{PC}), PacketTiming (\textsc{PT}), and TCPCounter (\textsc{TC}) feature
classes using subsets of our candidate feature set. \textsc{PC} includes all
packet and byte counters, \textsc{PT} includes all packet inter-arrival
statistics, and \textsc{TC} includes all flag counters, window size statistics,
and RTT. We simulate using Traffic Refinery by manually aggregating feature
classes, varying the packet depths, and measuring the predictive performance and
pipeline execution time using \sysname's Profiler. While Traffic Refinery also
has the ability to profile state and storage costs, we focus on execution time
in our evaluation since it is a shared cost metric in both frameworks.

\section{Simulated Annealing Details}
\label{app:simanneal_impl}

We describe our simulated annealing~\cite{kirkpatrick1983anneal}
implementation \simanneal from Section~\ref{sec:optimizer_eval}. \simanneal
starts with a random feature representation $x$ as the current ``best'' point.
On each iteration $i$, it samples a neighbor point $x_i$ and measures
$\texttt{cost}(x_i)$ and $\texttt{perf}(x_i)$. Neighbors are sampled by randomly
perturbing either the feature set or the packet depth with equal probability:
\begin{itemize}[noitemsep,topsep=0pt,leftmargin=15pt]
    \item Feature set perturbation: Add, remove, or replace a feature at random.
    \item Packet depth perturbation: Move up to some maximum step size away from
    the current packet depth, with the maximum step size decreasing linearly
    from the maximum packet depth as more samples are explored. This allows for
    more exploration earlier in the search.
\end{itemize}
Since our optimization is multi-objective, we adjust the standard simulated
annealing neighbor acceptance criterion as follows: If the neighboring point
dominates the current point across both objectives, it is accepted as the new
current point. Otherwise, it is still accepted with probability $\mathbb{P}(x,
x_i, T_i) = \exp{(f(x) - f(x_i))/T_i}$, where $f(x)$ is an equal weighted
combination of $\texttt{perf}(x)$ and $\texttt{cost}(x)$, and $T_i$ is the
temperature at iteration $i$. Consistent with simulated annealing algorithms,
the temperature gradually decreases as the search space is explored. This
mechanism allows for non-dominating solutions to still be accepted with higher
probability at the beginning of the search process, preventing \simanneal from
getting trapped in local minima. We empirically tune \simanneal with different
initial temperatures and cooling schedules, and choose $T_0 = 1$ and $T_{i+1} =
0.99T_{i}$.

Figure~\ref{fig:convergence} (Section~\ref{sec:optimizer_eval}) reveals that
while \simanneal is less sample efficient than \sysname, it is generally more
efficient than \randsearch. However, \randsearch catches up after approximately
1500~sample explorations, likely due to \simanneal's reduced ability to explore
new feature representations as the temperature decreases.

\begin{table}
    \centering
    \footnotesize
	\begin{tabular}{lrr@{}}
    \toprule
    \begin{tabular}[r]{@{}r@{}}\textbf{Use case / \# cand. features}\\
    \textbf{Systems cost metric:}\end{tabular} &
    \begin{tabular}[l]{@{}r@{}}\texttt{app-class} / 67\\ zero-loss
    throughput\end{tabular} & \begin{tabular}[c]{@{}r@{}}\texttt{iot-class}
    / 6\\ processing time\end{tabular} \\ \midrule \textbf{Preprocessing} &
    22.4~s & 4.1~s \\
    \textbf{Opt. Iteration} (50$\times$)&   & \\
    \hspace{1em} BO sample & 55.5~s & 1.4~s \\
    \hspace{1em} Pipeline generation & 53.1~s & 46.5~s \\
    \hspace{1em} Measure $\texttt{perf}(x)$ & 29.6~s & 26.4~s \\
    \hspace{1em} Measure $\texttt{cost}(x)$ & 546.7~s & 70.3~s \\
    \hline
    \textbf{Total elapsed} & 9.5~h & 2~h \\

    \bottomrule                          
\end{tabular}
	\vspace{-5pt}
	\caption{\sysname optimization
    wall-clock times. BO is well-suited for expensive-to-evaluate objective
    functions, such as $\texttt{perf}(x)$ and $\texttt{cost}(x)$.}
    \label{tab:wallclock}
	\vspace{-10pt}
\end{table}

\section{Ethical Considerations}
\label{app:ethics}
As part of our experiments with high-speed
network traffic, we evaluated the performance of models against live campus
network traffic.  The candidate flow features we use included only aggregate
flow statistics and an application name derived from the SNI field in TLS
handshakes. We never captured or analyzed client IP addresses, viewed any
individual flows or connection records, stored any packets to disk, or
investigated human behavior; our IRB has ruled that this type of analysis does
not constitute human subjects research.  Nonetheless, we took steps to ensure
the security and privacy of campus users. All live traffic experiments were
isolated to a single hardened server that was deployed in partnership with our
campus networking and security teams in order to not increase the attack surface
for users.\looseness=-1

\end{document}